\documentclass[10pt,twocolumn]{revtex4}
\usepackage{graphicx}
\usepackage{psfig}
\newcommand{\be}{\begin{equation}}
\newcommand{\bea}{\begin{eqnarray}}
\newcommand{\ee}{\end{equation}}
\newcommand{\eea}{\end{eqnarray}}
\newcommand{\eps}{\epsilon}

\newcommand{\dt}{\delta}
\newcommand{\pt}{\partial}

\begin{document}

\title{Universality in nonadiabatic behaviour of classical actions in nonlinear models with separatrix crossings.
}

\author{A.P. Itin,$^{1,2}$ S. Watanabe$^1$}
\affiliation{$^1$Department of Applied Physics and Chemistry,
University of Electro-Communications,\\ 1-5-1, Chofu-ga-oka,
Chofu-shi,
Tokyo 182-8585, Japan\\ $^2$Space Research Institute, Russian Academy of Sciences,\\
Profsoyuznaya str. 84/32, 117997 Moscow, Russia}\email{alx_it@yahoo.com}

\begin{abstract}

We discuss dynamics of approximate adiabatic invariants in several
nonlinear models being related to physics of Bose-Einstein
condensates (BEC). We show that nonadiabatic dynamics in Feshbach
resonance passage, nonlinear Landau-Zener (NLZ) tunnelling, and
BEC tunnelling oscillations in a double-well can be considered
within a unifying approach based on the theory of separatrix
crossings. The separatrix crossing theory was applied previously
to some problems of classical mechanics, plasma physics and
hydrodynamics, but has not been used in the rapidly growing
BEC-related field yet. We derive explicit formulas for the change
in the action in several models. Extensive numerical calculations
support the theory and demonstrate its universal character. We
also discovered a qualitatively new nonlinear phenomenon in a NLZ
model which we propose to call {\em separated adiabatic
tunnelling} (AT).
\end{abstract}
\maketitle

\section{Introduction}

Adiabatic invariance \cite{Ehrenfest} is very important in many
fields of physics.  In the last decade, there has been a great
deal of interest in physics of Bose-Einstein condensates
\cite{E1,E2,E3,E4,GP,Pit} among scientists from several scientific
fields. Presently BEC research is at the crossing point of AMO
science, statistical mechanics and condensed matter physics,
nonlinear dynamics and chaos. The discussion we present here is
related to interplay between nonlinearity and nonadiabaticity in
BEC systems. The relation between quantum transitions and change
in the classical action of a harmonic oscillator has long been
known \cite{Dykhne,Landau}. BEC bring nonlinearity into a quantum
world. BEC dynamics can often be described within the mean-field
approximation; finite-mode expansions produce nonlinear models
where a variety of phenomena common to classical nonlinear systems
happen. We consider two kinds of nonlinear phenomena here:
destruction of adiabatic invariance at separatrix crossings and
probabilistic captures in different domains of phase space.

A conceptual phenomenon of classical adiabatic theory is
destruction of adiabatic invariance at separatrix crossings which
is encountered, in particular, in plasma physics and
hydrodynamics, classical and celestial mechanics
\cite{AKN,T,Cary,N86,N87,Skodje,INV00,NV99,INV01,Bil,Chaos,Dietz,Zaslavsky}.
The phenomenon is very important for BEC physics: we consider here
nonlinear two-mode models related to tunnelling between coupled
BEC in a double well \cite{Smerzi}, nonlinear Landau-Zener
tunnelling \cite{Zobay,Niu}, Feshbach resonance passage in
atom-molecule systems \cite{Vardi,Comment,PhysD}. Nonlinear
two-mode models were extensively studied previously (sometimes
beyond the mean-field approximation, \cite{Smerzi,Zobay,Niu,
Sacha, Abdullaev2000,
Elyutin,Kivshar,Reinhardt,Meystre,Yurovsky,Milburn,Adhikari,Bergman,
Javanainen,Javanainen2,Timmermans,Levitov,Links,Altshuller}), and
destruction of adiabaticity was discussed already in
\cite{Zobay,Niu,Vardi}, still  there are regimes of motion that
were not analyzed in these papers from the point of view of
nonadiabatic behaviour, that is, when initial populations of both
modes are not zero (or very small), but finite. We presented some
of our results on that theme in \cite{Comment,PhysD}; nevertheless
destruction of adiabatic invariance has not been studied
systematically in BEC-related models yet. Action is an approximate
adiabatic invariant in a classical Hamiltonian system that depends
on a slowly varying parameter provided a phase trajectory stays
away from separatrices of the unperturbed (frozen at a certain
parameter value) system. If this condition is not met,
adiabaticity may be destroyed. As the parameter varies, the
separatrices slowly evolve on the phase portrait. A phase
trajectory of the exact system may come close to the separatrix
and cross it. The general theory of the adiabatic separatrix
crossings \cite{AKN} predicts universal behavior of the classical
action (described in a greater detail in the main text). In
particular, at the separatrix crossing the action undergoes a
quasi-random {\em dynamical} jump, which is very sensitive to
initial conditions and depends on the rate of change of the
parameter. The asymptotic formula for this jump  was obtained in
\cite{T,Cary,N86}. Later, the general theory of adiabatic
separatrix crossings was also developed for slow-fast Hamiltonian
systems \cite{AKN,NV99}, and was applied to certain physical
problems (see, for example, \cite{INV00,INV01,Bil,Chaos}). It was
also noticed that nonlinear Landau-Zener (NLZ) tunnelling models
constitute a particular case for which the general theory can be
applied \cite{Comment,PhysD}.
 Beside the quasi-random jumps of adiabatic invariants, there is
another important mechanism of stochastization in BEC-related
models: scattering on an unstable fixed point with a capture into
different regions of phase space after a separatrix crossing
\cite{Slutskin,N75,Four}. Here stochastization happens due to
quasi-random splitting of phase flow in different regions of phase
space at the crossing. Rigorous definition of such probabilistic
phenomena in dynamical systems was given in \cite{Arnold63}. The
probabilistic capture is important in problems of celestial
mechanics \cite{AKN}, but it was also investigated in some
problems of plasma physics and hydrodynamics  \cite{INV00}, optics
\cite{INV01}, classical billiards with slowly changing parameters
and other classical models \cite{Bil}. As shown in \cite{Four},
the combination of the two phenomena leads to {\em dephasing} in
dynamics of globally coupled oscillators modelling coupled
Josephson junctions. However, it seems that the probabilistic
capture mechanism was not discussed at all in relation to BEC
models yet. We discovered that in a nonlinear Landau-Zener model
such mechanism may take place, and it leads to a new phenomenon
(in the context of the model) that we propose to call {\em
separated adiabatic tunnelling}.

Let us review the models being considered in the present paper in
more concrete terms. The nonlinear two-mode model introduced in
\cite{Smerzi} describes BEC tunnelling oscillations in a
double-well as that of a non-rigid pendulum. In the case of an
asymmetric double-well, the effective classical Hamiltonian is:
\be H=-\delta w +\frac{\lambda w^2}{2} -\sqrt{1-w^2} \cos \theta,
\label{ham2ABEC} \ee where $w,\theta$ are the population imbalance
and phase difference between the two modes, parameters $\delta$
and $\lambda$ represent potential difference between the wells and
nonlinearity, correspondingly. The same Hamiltonian appears in a
nonlinear Landau-Zener model \cite{Zobay,Niu}. As one slowly
sweeps $\delta$, say, from a large positive to a large negative
value, a change in the mode populations is determined by the
change in the classical action (since at large $|\delta|$
classical action depends linearly on $w$). This provides
interesting link between fundamental issue of classical mechanics,
dynamics of approximate adiabatic invariants (classical actions),
and nonadiabatic transitions in quantum many-body systems. The
dynamics of classical actions in nonlinear systems is, however, a
very complicated issue \cite{AKN}. Some analysis of the NLZ model
was done in \cite{Niu,Zobay}. In \cite{Niu} so-called subcritical
($\lambda<1$), critical ($\lambda=1$),  and supercritical
($\lambda>1$) cases were defined. However, only the case of zero
initial action was considered, that is a vanishingly small initial
population in one of the states. We concentrate on the case of
finite initial action, and supercritical case. In the
supercritical case, the most striking phenomenon from the point of
view of physics is the so-called {\em nonzero adiabatic
tunnelling} (nonzero AT). In terms of the theory of separatrix
crossings, it is caused by the {\em geometric} jump in the action
at the separatrix crossing. Mathematically, it is a very simple
issue: as a phase point leaves a domain bounded by a separatrix of
the unperturbed system and enters another domain, its action
undergoes a ''geometric" change proportional to the difference in
areas of the two domains \cite{dynamical}. The gist of {\em
separated adiabatic tunnelling} (seoarated AT) that we found is as
follows. The separatrix divides the phase portrait on 3 domains
$G_{1,2,3}$ (Fig. (\ref{FNLZ})). In case at the moment of
separatrix crossing areas of $G_{1,2}$ grow, the phase point
leaving the third domain $G_3$ (with decreasing area) can be
captured in either of the two growing domains. Bunch of
trajectories with close initial actions $I_i$ will be "splitted"
on two bunches with two different final actions $I^{1,2}_f$. It is
possible to calculate probability for a phase point to come to
either of the two bunches (we calculated it in Appendix B and
compared analytical prediction with numerical result in Fig. 11).
Possible physical applications of nonzero AT phenomenon has been
extensively discussed (for example, \cite{Niu,Zobay}: wavepacket
in an accelerated optical lattice should undergo nonzero
tunnelling in adiabatic limit when nonlinearity is large enough,
although no experimental evidence is available yet). In relation
to BEC oscillations in asymmetric double-well, corresponding
physical effect is (obvious) drastic change in the amplitude of
oscillations when regime of motion is changed from self-trapped to
complete oscillations due to slow change of parameters. In the
case of {\em separated AT}, the effect for the asymmetric well may
look like this: the asymmetry between the wells is slowly changed;
regime of motion is changed from self-trapped to complete
oscillations and then back to self-trapped. But final state is
"splitted": a system has "choice" of two different final states.
For a set of experimental realizations with close initial
conditions, one can define a "probability" for a system to come to
either of the two states. While such experimental realization
seems to be even less realistic than nonzero AT, conceptually it
is a very interesting phenomenon worth discussing: the
"probability" is of purely classical origin. Analogous
interpretation can be done for BEC experiencing NLZ tunnelling in
optical lattice. Although the phenomenon looks similar to the
nonzero AT described in \cite{Niu,Zobay}, its mathematical
background is very much different and not so straightforward; it
is a particular case of probabilistic phenomena in dynamical
systems defined in \cite{Arnold63}.

We also derive a formula for the jump of the adiabatic invariant
(Eq. (\ref{symmjump}) ) in the symmetric well case ($\delta=0$)
and check it numerically (for the asymmetric case, the
corresponding formula has both terms of order $\eps$ and $\eps \ln
\eps$). As physical application of this jump, one can imagine an
experiment with BEC oscillations in a double-well, with the
potential barrier between the wells being slowly raised and then
slowly decreased back to its initial position. The system will not
return back to its initial state. Within the mean-field two-state
model, the difference between the initial and final oscillations
is caused by the change in the adiabatic invariant (of course, in
a real system many other complications arise). Such kind of
experiments are feasible \cite{MO,Albiez}.

Similar nonadiabatic phenomena arise in coupled atom-molecular
systems. Here, in the mean-field limit it is possible to construct
two-mode models based on the all-atom and all-molecule modes, and
their coherent superpositions.
\begin{figure*}
\includegraphics[width=160mm]{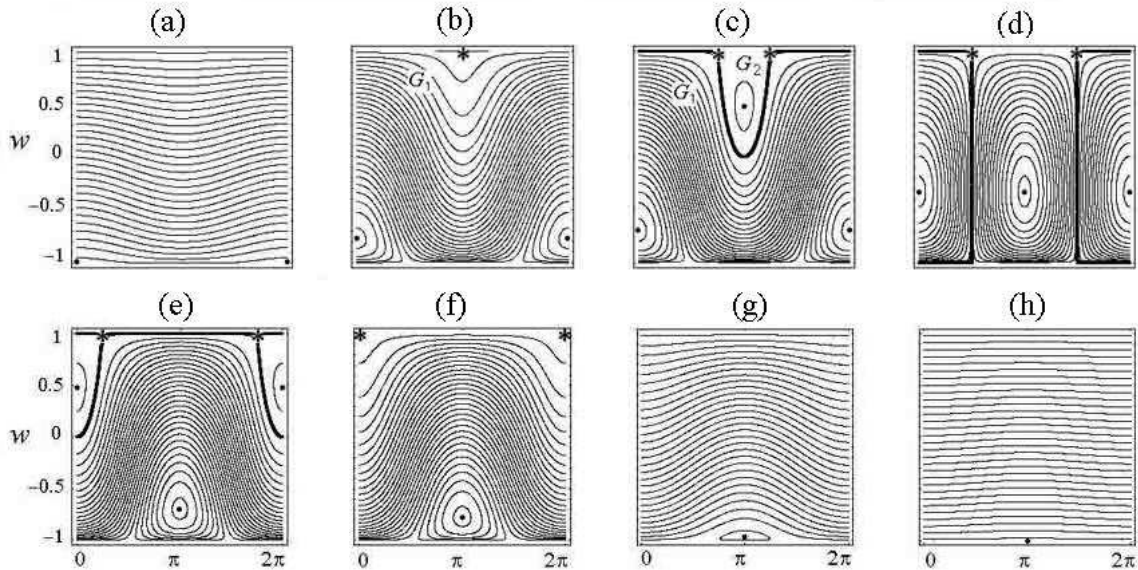}
\caption{Phase portraits of the Hamiltonian (\ref{ham2modem}) with
$\lambda=0$. From left to right: $\delta=10,
\sqrt{2},1,0,-1,-\sqrt{2},-5,-50$. Stars (bold dots): unstable
(stable) fixed points. See detailed discussion in
\cite{Comment,PhysD} \label{fermi} }
\end{figure*}
The two-mode model describing a degenerate gas of fermionic atoms
coupled to bosonic molecules was considered in
\cite{Vardi,Comment,PhysD} (the same model enables to describe
coupled atomic and molecular BECs, so we call it 2-mode AMBEC
model). The system is reduced to the classical Hamiltonian \be H =
-\dt(\tau) w + (1-w)\sqrt{1+w} \cos \theta, \label{ham} \ee where
$w$ denote population imbalance between atomic and molecular
modes, and $\delta$ is the (slowly changing) detuning from the
Feshbach resonance. As $\delta$ sweeps from large positive to
negative values, the system is transferred from the all-atom $w=1$
mode to the all-molecule $w=-1$ mode. The final state of the
system contains the non-zero remnant fraction, which can be
calculated as change in the classical action in the model
(\ref{ham}), and scales as a power-law of the sweeping rate. The
model was introduced in \cite{Vardi} in the attempt to describe
recent experiments on Feshbach resonance passage
\cite{Regal,Hulet,Cubizolles,Hogby}, and some power laws were
calculated there and compared with experimental data. For the case
of nonzero initial molecular fraction, the power-law was also
calculated in \cite{Comment,PhysD} according to the general
theory. We carefully check numerically this (linear) power law in
Section IIB. It is important to note that the model give 100\%
conversion efficiency in the adiabatic limit, while in the
experiments finite conversion efficiency has been seen. In Section
IIC we present brief analysis of a more general model, which have
an analog of nonzero AT (leading to finite conversion efficiency
in the adiabatic limit). In the more general version, s-wave
interactions were taken into account, so the Hamiltonian looks
like

\be H =-\delta w + \lambda w^2 + (1-w)\sqrt{1+w} \cos \theta,
\label{2AMBEC} \ee

Here, the phase portraits can have more complicated structure, and
the passage through the separatrix can be accompanied by the
geometric jump in the action, leading to a non-zero remnant
fraction even in the adiabatic limit.

In Section III, the nonlinear two-mode model (\ref{ham2ABEC}) for
two coupled BECs is considered. For brevity, we call this model
2-mode atomic BEC (ABEC) model.  The {\em separated AT} is
demonstrated in the end of the Section.
\begin{figure*}
\includegraphics[width=140mm]{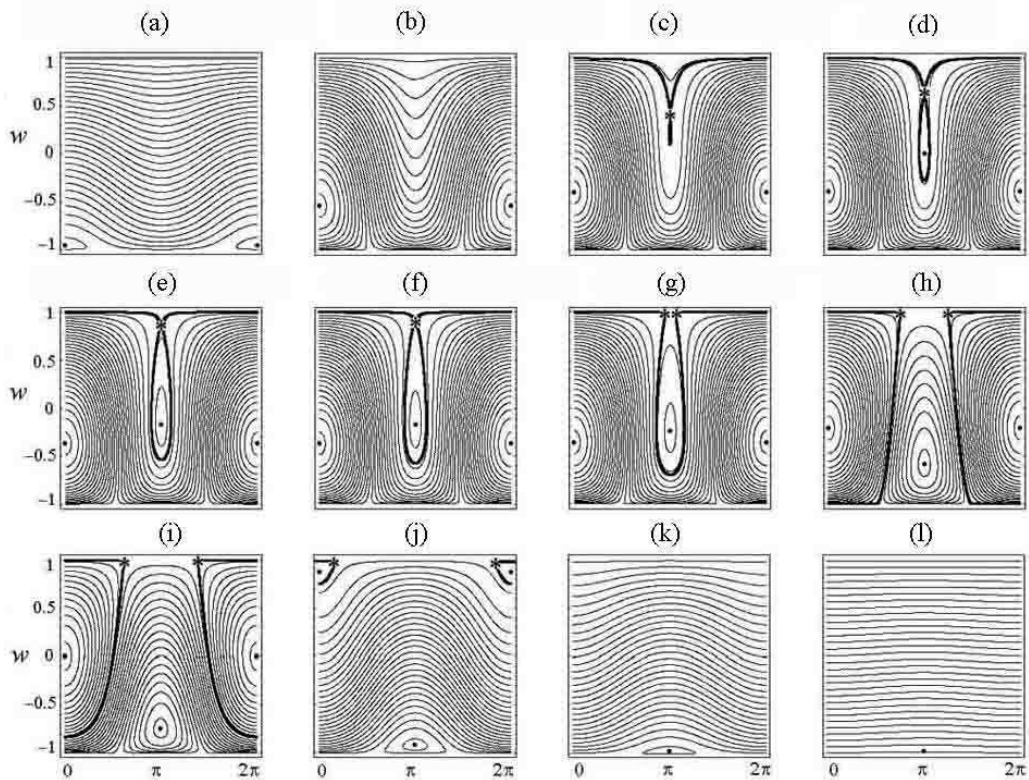}
\caption{Phase portraits of the Hamiltonian (\ref{ham2modem}) with
 $\lambda<0$ $(\lambda=-0.5)$. From upper left to bottom right (a-l): $\delta=
 5.0,1.0,0.53,0.5,0.45,0.44,0.4,0,-0.5,-2.2,-5,-50$. In (c)-(f), the separatrix divide phase portraits on three domains $G_{1,2,3}$ ($G_2$ is adjacent to the segment $w=1$, $G_1$ is adjacent to $w=-1$, $G_3$ is the loop in between).
 Starting with small initial action at $w \sim 1$, a phase point undergoes
 a geometric jump in the action in addition to a dynamical jump. This leads to analog of nonzero AT and finite conversion
efficiency in the adiabatic limit. \label{AMBECportrait}}
\end{figure*}

\begin{figure}
\includegraphics[width=50mm]
{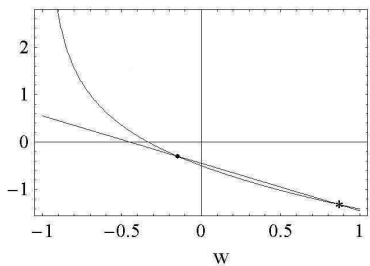} \caption{Graphical solution of the equation (\ref{FP}). The
line $y(w)=2 \lambda w - \delta$ crosses the curve $y(w)=-\frac{3
w +1 }{2 \sqrt{w+1}}$ in two points (provided $2 \lambda <
\mbox{max} \{y'(w)\}=-1/\sqrt{2}$), one of the points corresponds
to the unstable fixed point on the phase portraits of Figs.2c-2f,
while the other to the stable elliptic point at $\theta=\pi$ in
these Figs. As $\delta$ decreases further, the unstable fixed
point moves to $w=1$.}
\end{figure}
Section IV contains concluding remarks. In the Appendix A we
described the adiabatic and improved adiabatic approximations. In
the Appendix B, derivation of formula for probabilities in
separated AT is presented.

The most interesting new results of the paper are:

1) Extensive numerical tests of the formula (\ref{jump}) for the
dynamical jump in the adiabatic invariant of the atom-molecule
system (Section IIb). The formula is based on Eq.(\ref{jumpp})
being obtained elsewhere \cite{Comment,PhysD}.

2) Suggested mechanism (analog of nonzero AT, Section IIc, Fig.
\ref{AMBECportrait}) leading to finite conversion efficiency in
atom-molecule systems due to a geometric jump in the action.

3) Analytical derivation of the explicit expression
(\ref{symmjump}) for the jump of adiabatic invariant of the
symmetric ABEC model and its numerical test (Section IIIb,
Fig.\ref{symmjumpcheck}).

4) Discovery of new phenomenon in nonlinear Landau-Zener model:
{\em separated AT}. Analytical calculation of probabilities
related to this tunnelling (B32-34) and its numerical test
(Section IIIc, Fig. \ref{fcapture}).

The {\em main} result is demonstration of usefulness of separatrix
crossing theory in a variety of BEC-related models.

In order to keep the paper compact, we do not present here
comparison with quantum many-body calculations, but consider only
mean-field models.

\section{Nonlinear two-mode models for atom-molecular systems.}
\subsection{Model equations and its physical origin; classical phase portraits}

\begin{figure*}
\includegraphics[width=140mm]
{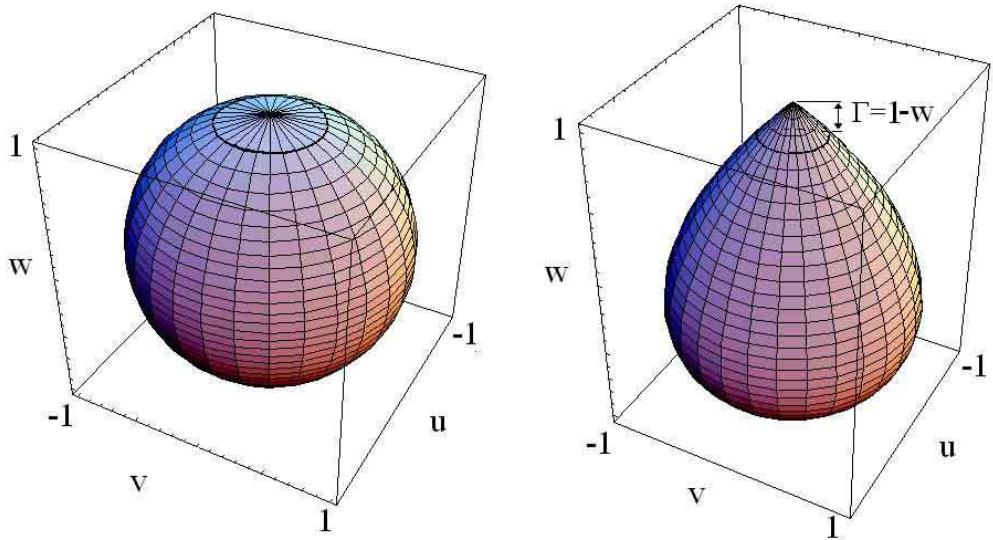} \caption{The Bloch sphere corresponding to ABEC models and
the generalized Bloch sphere corresponding to AMBEC models (the
surfaces $u^2+v^2=1-w^2$ on the left and
$u^2+v^2=\frac{1}{2}(w-1)^2(w+1)$ on the right). At large
detuning, near $w=1$, the area within a trajectory on the
generalized Bloch sphere is proportional to $u^2+v^2 \approx
(1-w)^2 =\Gamma^2$, while on Bloch sphere the area is proportional
to $u^2+v^2 \approx 2 (1-w) = 2 \Gamma$ . Note however that action
variable in either case is proportional to $1-w$. Action is
related to the area on the Hamiltonian phase portraits which is
approximately equal to $1-w$ for the corresponding trajectory, see
 \cite{Comment,PhysD}. \label{Bloch}}
\end{figure*}

In BEC-related mean-field models nonlinearity usually comes from
s-wave interactions (via a scattering length entering the
nonlinear term of the Gross-Pitaevskii equation \cite{GP}).
However, interesting nonlinear models arise in atom-molecular
systems, where atoms can be converted to BEC of molecules. Even
neglecting collisions and corresponding s-wave interactions, the
nonlinearity comes into play from the fact that two atoms are
needed to form a molecule.

We consider a Hamiltonian system with the Hamiltonian function \be
H = -\dt(\tau) w + \lambda w^2 + (1-w)\sqrt{1+w} \cos \theta,
\label{ham2modem} \ee where $\tau=\eps t,$ $\eps \ll 1$. Several
systems can be described by the model (\ref{ham2modem}), in
particular coupled atomic and molecular BEC \cite{Yurovsky}, and a
gas of fermionic atoms coupled to molecular BEC
\cite{Vardi,Comment,PhysD}. Let us briefly discuss these systems.

In \cite{Yurovsky}, a system of coupled atomic and molecular
condensates was considered using generalization of the Bloch
representation for the two-mode system. Quantum Hamiltonian of the
system is $ \hat H = \frac{\Delta}{2} a ^{\dagger} a +
\frac{\Omega}{2} ( a ^{\dagger} a^{ \dagger} b + b^{ \dagger }a a
) $, where $a ^{\dagger}$ and $a$ are the creation and
annihilation operators of the atomic mode, while $b^{ \dagger }$
and $b$ are the creation and annihilation operators of the
molecular mode. The two modes are supposed to be coupled by means
of a near resonant two-photon transition or a Feshbach resonance,
with a coupling frequency $\Omega$ and detuning $\Delta$.
Introducing operators $\hat L_x = \sqrt{2} \frac{ a ^{\dagger} a^{
\dagger} b + b^{ \dagger }a a }{N^{3/2}} $, $\hat L_y = \sqrt{2}
\frac{ a ^{\dagger} a^{ \dagger} b - b^{ \dagger }a a }{i
N^{3/2}}$, $ L_z = \frac{2 b^{ \dagger } b -   a ^{\dagger} a
}{N}$,  Heisenberg equations of motion in the mean-field limit
lead to the dynamical system for the rescaled components of
generalized Bloch vector:  $ \dot{s}_x = - \delta s_y, \quad
\dot{s}_y = -\frac{\sqrt{2}}{4}(s_z-1)(3 s_z+1) + \delta s_x,
\quad \dot{s}_z=-\sqrt{2} s_y, $ where the rescaled detuning is
$\delta=\Delta/(\sqrt{N} \Omega)$, while $s_{x,y,z}$ are the
expectation values of $L_{x,y,z}$ ($s_z=1$ corresponds to
all-molecule mode;  $N$ is the number of atoms). Exactly the same
dynamical system arises in the degenerate model of fermionic atoms
coupled to BEC of diatomic molecules \cite{Vardi}. Indeed, using a
similar approach in \cite{Vardi} an analogous system of equations
was obtained
 \bea
\dot u &=& \dt(\tau)v, \nonumber \\
\dot v &=& -\dt(\tau)u + \frac{\sqrt 2}{4} (w-1)(3w+1),
\label{uvsystem}\\
\dot w &=& \sqrt 2 v, \nonumber \eea where $w$ is the population
imbalance between all-atom and all-molecule mode, $u$ and $v$ are
real and imaginary parts of the atom-molecule coherence, $\delta$
is the rescaled detuning from the resonance. These equations are
equivalent to the Hamiltonian equations of motion of the
Hamiltonian system (\ref{ham2modem}) with $\lambda=0$
\cite{Comment}. The variable $\theta$ canonically conjugated to
$w$ is related to the old variables as $\theta=\mbox{atan}(v/u)$.
The all-atom mode now corresponds to $w=1$, while all-molecule
mode to $w=-1$. Sweeping through Feshbach resonance from fermionic
atoms to bose molecules can be described by the Hamiltonian
(\ref{ham2modem}) with $\lambda=0$ and $\delta$ slowly changing
from large positive to large negative values. In both systems
(atom-molecule BEC and degenerate fermionic gas coupled to BEC of
molecules) mean-field collisional interactions were neglected so
far. The case $\lambda \ne 0$ in the Hamiltonian (\ref{ham2modem})
corresponds to inclusion of the s-wave scattering interactions.
Recently, in \cite{Links} a  more general quantum Hamiltonian
describing the coupling between atomic and diatomic-molecular BECs
within two-mode approximation was considered: \bea H &=& U_a N_a^2
+ U_b N_b^2 + U_{ab} N_aN_b + \mu_a N_a + \nonumber\\ \mu_b N_b
&+& \Omega ( a ^{\dagger} a^{ \dagger} b + b^{ \dagger }a a), \eea
where $a^{\dagger}$ is the creation operator for an atomic mode
while $b^{ \dagger}$ creates a molecular mode; parameters $U_i$
describe S-wave scattering: atom-atom ($U_a$), atom-molecule
($U_{ab}$), and molecule-molecule ($U_b$). The parameters $\mu_i$
are external potentials and $\Omega$ is amplitude for the
interconvertions of atoms and molecules. $N_a$ and $N_b$ are
populations of the atomic and molecular mode, correspondingly. In
the limit of large $N=N_a + 2 N_b$, the classical Hamiltonian was
obtained: \be H= \lambda z^2 + 2 \alpha z + \beta + 2 \sqrt{1-z}
(1+z) \cos (4\theta/N), \label{Links} \ee where

\bea \lambda= \frac{\sqrt{2N}}{\Omega}(U_a/2-U_{ab}/4+U_b/8), \nonumber\\
\alpha =\frac{\sqrt{2N}}{\Omega}( U_a/2-U_b/8+\mu_a/2N -\mu_b/4N),
 \eea
$\theta$ is phase difference between the modes, $z$ is difference
in populations.
 It is not difficult to transform the Hamiltonian (\ref{Links})
to the form (\ref{ham2modem}) denoting $z=-w$ and introducing a
new time variable $t'=4t/N $ to get rid of the $4/N$ multiplier in
the last term of (\ref{Links}). The term $\beta$ is not important
for dynamics. Therefore, the Hamiltonian (\ref{ham2modem})
describes coupled atomic-molecular BECs (with s-wave interactions)
in the mean-field limit. Sweeping through Feshbach resonance can
be modelled now by changing $\delta$ and keeping $\lambda$ fixed
in the Hamiltonian (\ref{ham2modem}). Self-trapping phenomenon in
the model discussed in \cite{Links} allows to predict
qualitatively new effect, that is non-zero remnant fraction in the
adiabatic passage through the resonance; we do not present
detailed quantitative analysis of the model in the present paper,
but note that it may provide an alternative explanation of finite
conversion efficiency at Feshbach resonance passage within
mean-field approximation. Similar to approach of \cite{Vardi}
mentioned above, s-wave interactions within molecular BEC can be
included in the model of fermionic atoms-bose molecules system via
the same coefficient $\lambda \ne 0$ in (\ref{ham2modem}).

\begin{figure}
\includegraphics[width=60mm]
{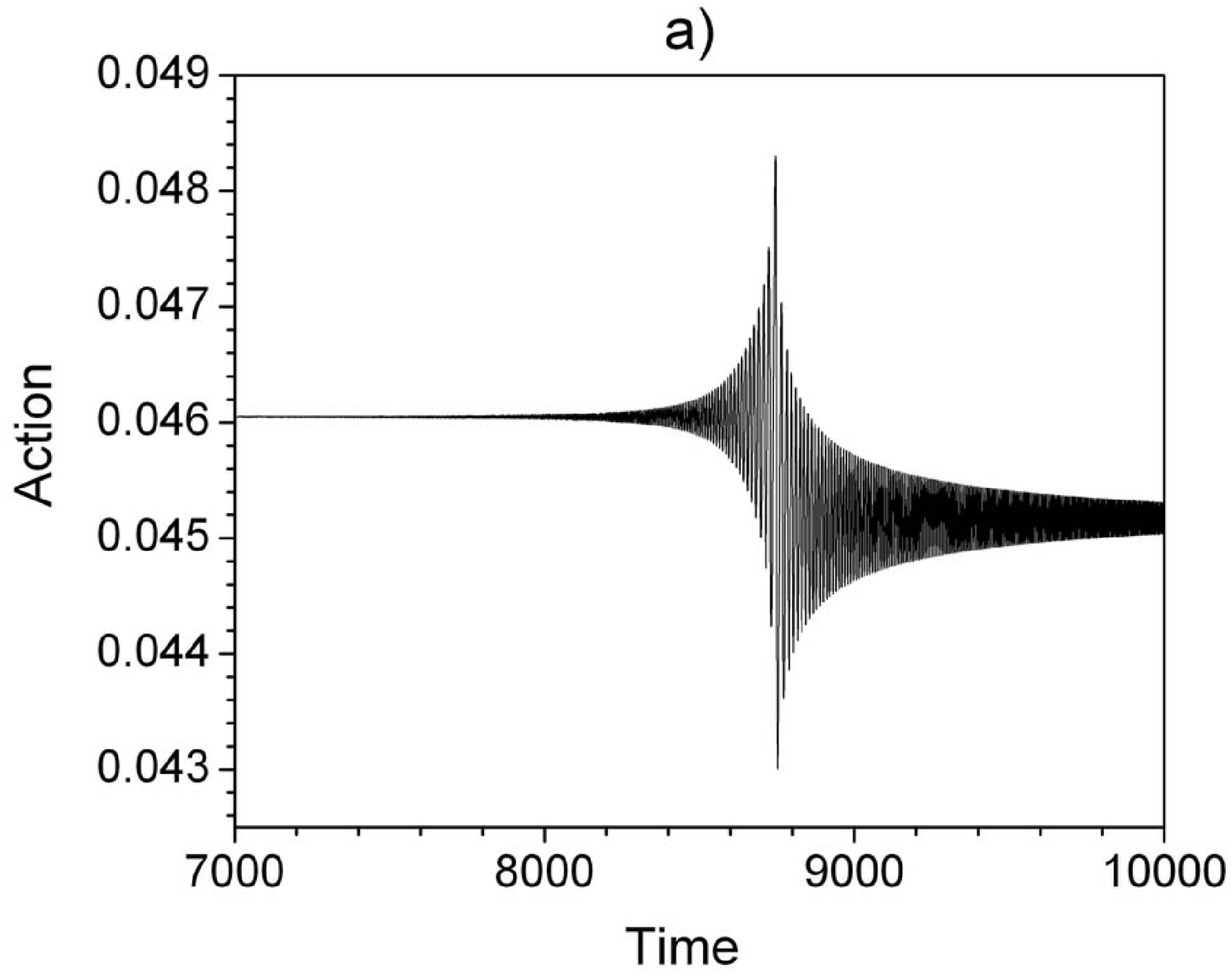}
\includegraphics[width=60mm]
{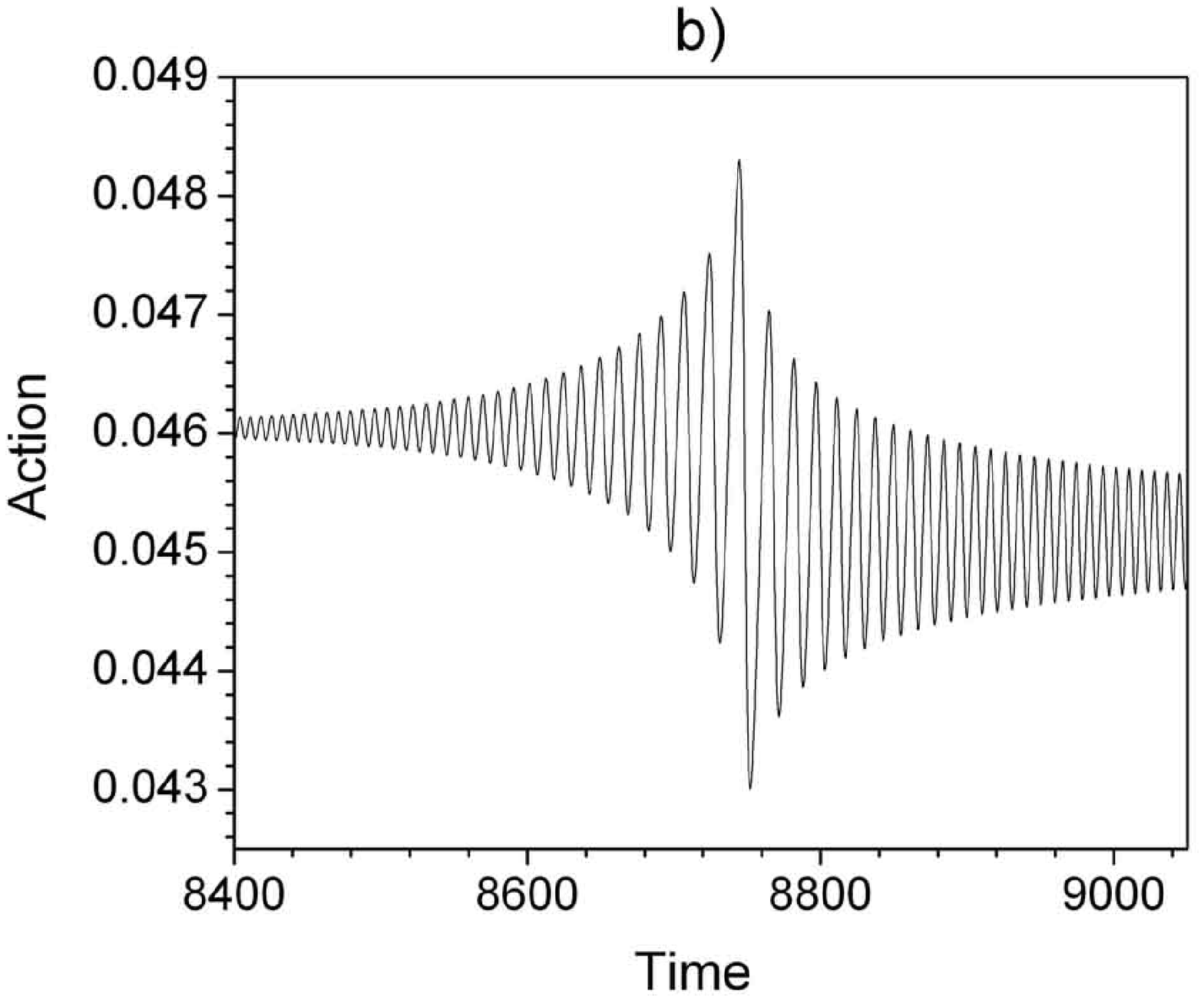}
\includegraphics[width=60mm]{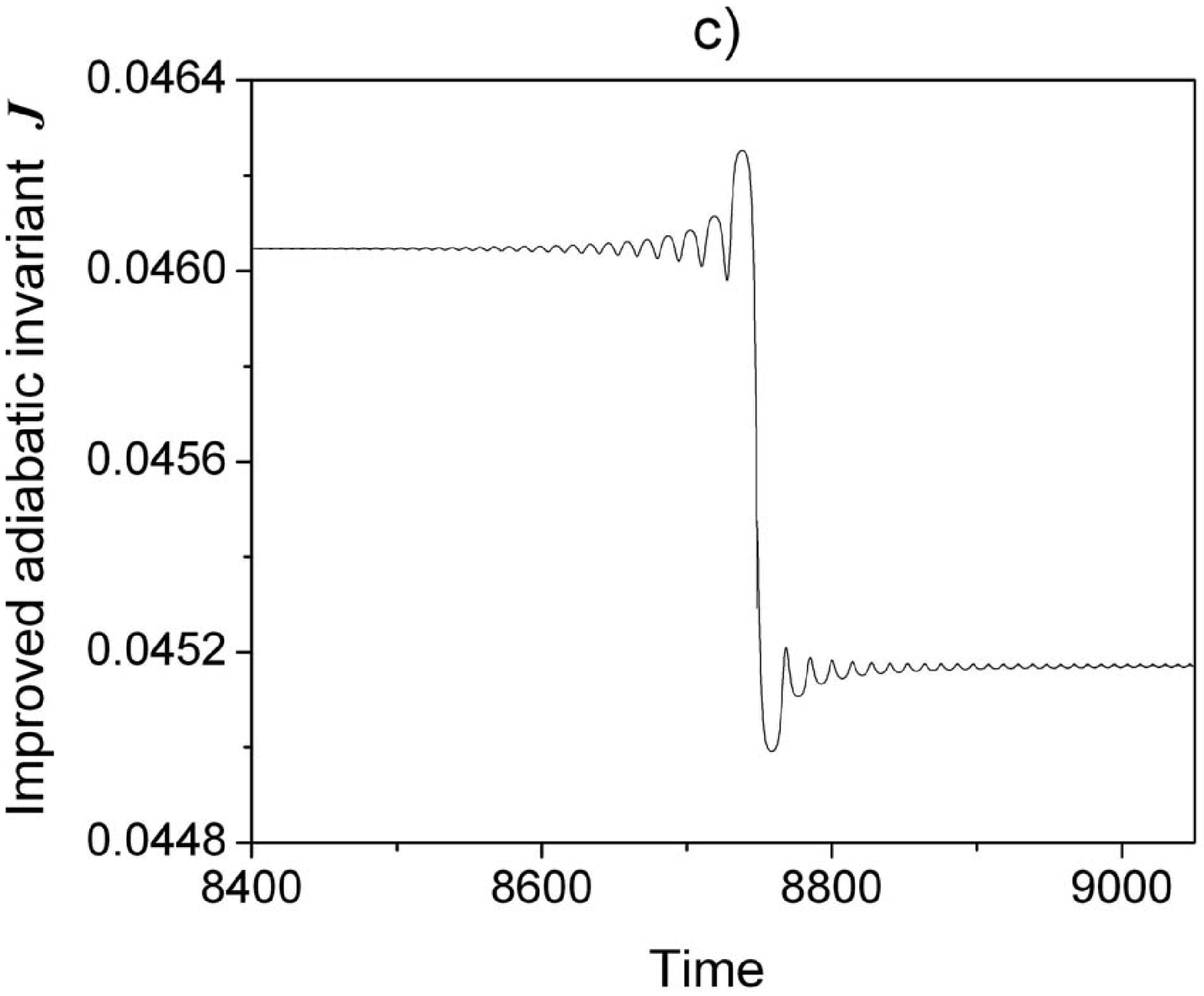}
\caption{Time evolution of the adiabatic invariant (action) $I$
and the improved adiabatic invariant $J$ in the model
(\ref{ham2modem}) with $\lambda=0$. \label{singlejump}}
\end{figure}

Phase portraits with $\lambda=0$ (Case I) and different values of
$\delta$ are given at Fig. \ref{fermi}. Phase portraits with some
constant $\lambda<0$ (Case II) and different values of $\delta$
are given at Fig.\ref{AMBECportrait}. The phase portraits for Case
I were analyzed in detail in  \cite{Comment}. The dynamics can
also be visualized using variables $u,v,w$ of the system
(\ref{uvsystem}). The latter system possesses an integral of
motion $u^2+v^2 - \frac{1}{2}(w-1)^2(w+1)=0$ defining the
generalized Bloch sphere (see Fig.4). The important property of
the generalized Bloch sphere is the singular (conical) point at
(0,0,1). As described in  \cite{Comment}, the  points $(0,0,\pm
1)$ are represented by the segments $w=\pm 1$ in the Hamiltonian
phase portraits. Nevertheless, it does not mean that all the
points of the either segment are equivalent. As described in
\cite{Comment,PhysD}, saddle points appear on the segment $w=1$ at
certain values of the parameter $\delta$. This drastically
influence dynamics in the vicinity of $w=1$. Let us briefly recall
the description of the phase portraits previously given in
\cite{Comment}.

If $\delta > \sqrt 2$, there is only one stable elliptic point on
the phase portrait, at $\theta = 0$ and $w$ not far from $-1$ [see
Figure 1a]. At $\delta = \sqrt 2$ a bifurcation takes place, and
at $\sqrt{2}>\delta> 0$ the phase portrait looks as shown in
Figure 1c. There are two saddle points at $w = 1,
\cos\theta=-\dt/\sqrt 2$ and a newborn elliptic point at $\theta =
\pi$. The trajectory connecting these two saddles separates
rotations and oscillating motions and we call it the separatrix of
the frozen system (what is most important is that the period of
motion along this trajectory is equal to infinity). At $\delta =
0$ on the phase portrait the segment $w=-1$ belongs to the
separatrix (Fig. \ref{AMBECportrait}d). At $0<\dt< \sqrt 2$ the
phase portrait looks as shown in Fig. \ref{AMBECportrait}e. At
$\dt=-\sqrt 2$ the bifurcation happens, and finally, at large
positive values of $\dt$, again there is only one elliptic
stationary point at $\theta = \pi$, and $w$ close to $-1$.

Let us introduce the action variable. Consider a phase trajectory
on a phase portrait frozen at a certain value of $\delta$. If the
trajectory is closed, the area $S$ enclosed by it is connected
with the action $I$ of the system by a simple relation $S = 2\pi
I$. If the trajectory is not closed, we define the action as
follows. If the area $S$ bounded by the trajectory and lines $w=1,
\theta=0, \theta=2\pi$ is smaller than $2\pi$, we still have $S =
2\pi I$. If $S$ is larger than $2\pi$, we put $2\pi I = 4\pi - S$.
Defined in this way, $I$ is a continuous function of the
coordinates.

How does the process of Feshbach resonance passage happen in terms
of the classical portraits of Fig.\ref{fermi}? Suppose one starts
with $w(0)=w_0 \approx 1$, and $\delta(0) \gg 1$ (physically, it
means that almost all population is in the atomic mode, but there
is small initial molecular fraction). In the phase portrait of the
unperturbed system the corresponding trajectory looks like a
straight line (Fig. \ref{fermi}a). The initial action of the
system approximately equals to $1-w_0$. For example, assume that
the area $S_*$ within the separatrix loop in
 Fig. \ref{fermi}c (corresponding to $\delta$=$\delta_*=1$) is
equal to $S_* = 2\pi I_0=2\pi (1-w_0)$. When, as $\delta$ slowly
decreases, the trajectory on an unperturbed phase portrait
corresponding to the exact instantaneous position of the phase
point $\{w(t),\theta(t) \}$ slowly deforms, but the area bounded
by it remains approximately constant: action is the approximate
adiabatic invariant far from the separatrix \cite{AKN}. As
$\delta$ tends to $\delta_*$, the form of the trajectory tends to
the form of the separatrix loop in Fig. \ref{fermi}c. The phase
point is forced to pass near the saddle point at the $w=1$ segment
many times. Since the area $S$ within the separatrix loop slowly
grows, approximately at the moment $\tau=\tau_*$ when $
\delta(\tau_*)=\delta_* $ separatrix crossing occurs, and the
phase point changes its regime of motion from rotational to the
oscillatory around the elliptic point inside the separatrix loop.
Then, it follows this elliptic point adiabatically (as no
separatrix crossings occur anymore). The elliptic point reaches
$w=-1$ at large positive $\delta$. The value of the population
imbalance tends to some final value $w=w_f$. The action variable
at large $\delta$ is approximately equal to $1+w$. We see that in
the adiabatic limit the sign of the population imbalance is
reversed, $w_0=-w_f$. Nonadiabatic correction to this result arise
due to the separatrix crossing and is discussed in detail in the
next paragraph.

\begin{figure*}
\includegraphics[width=80mm]
{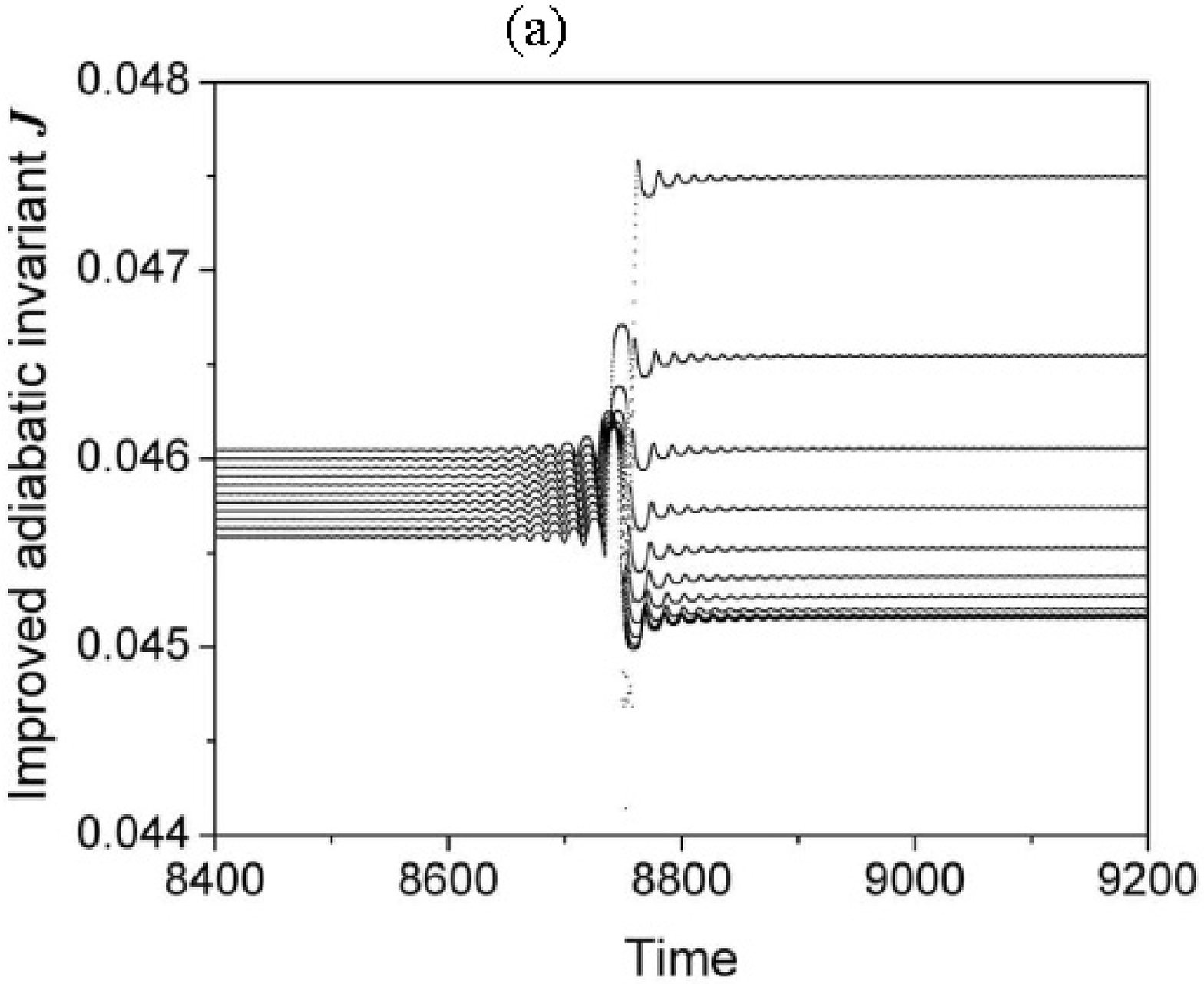}
\includegraphics[width=80mm]
{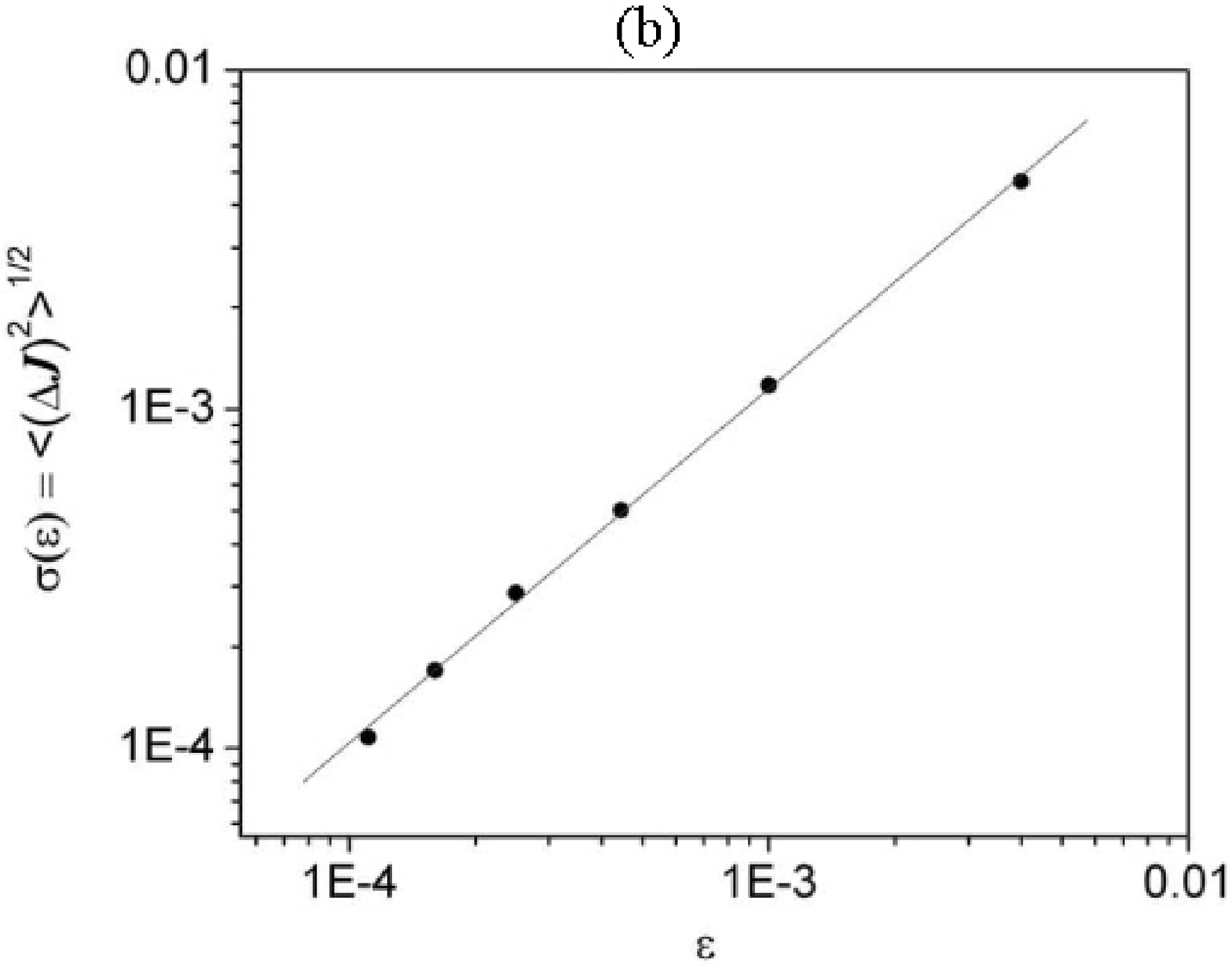}
\includegraphics[width=80mm]
{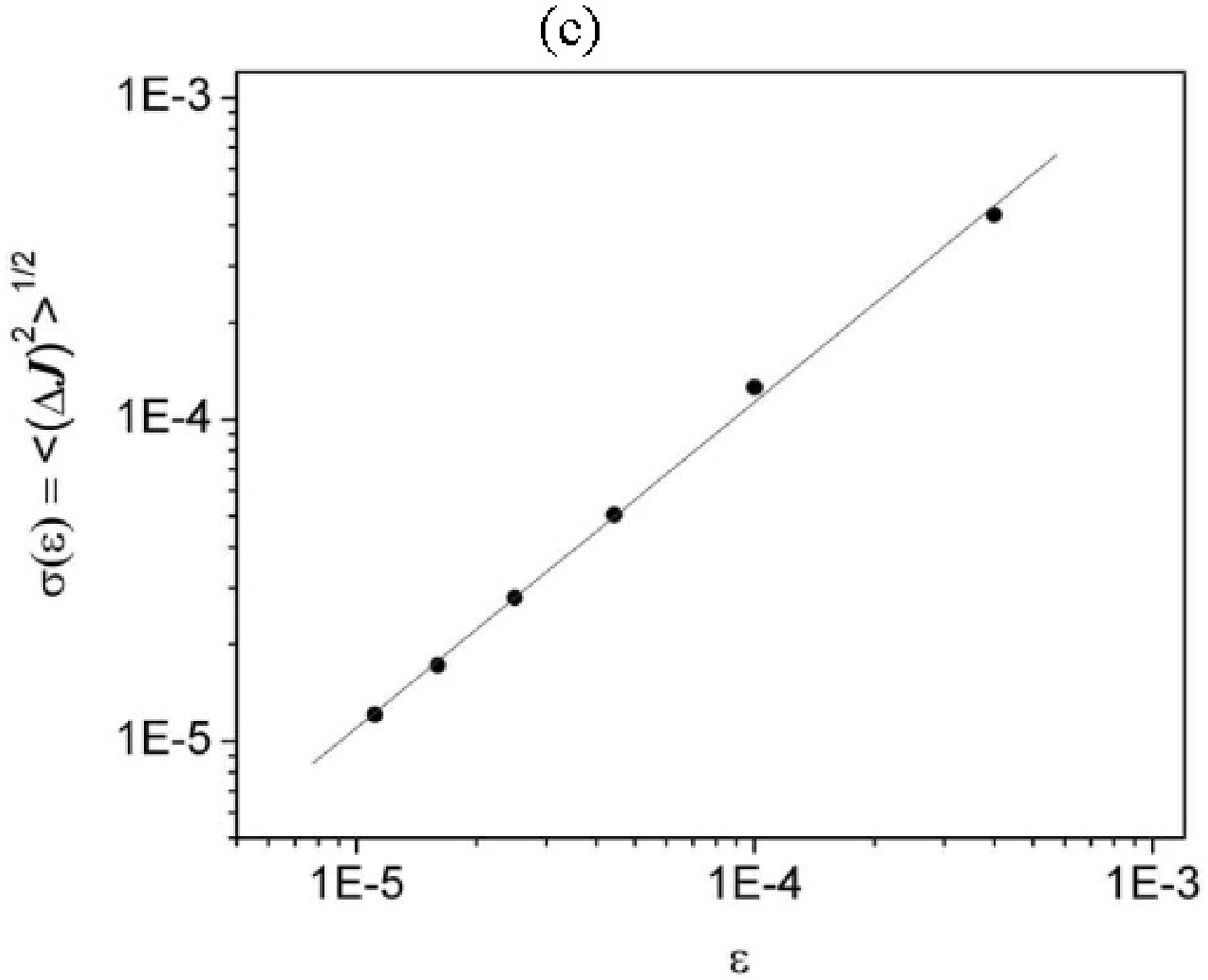}
\includegraphics[width=80mm]
{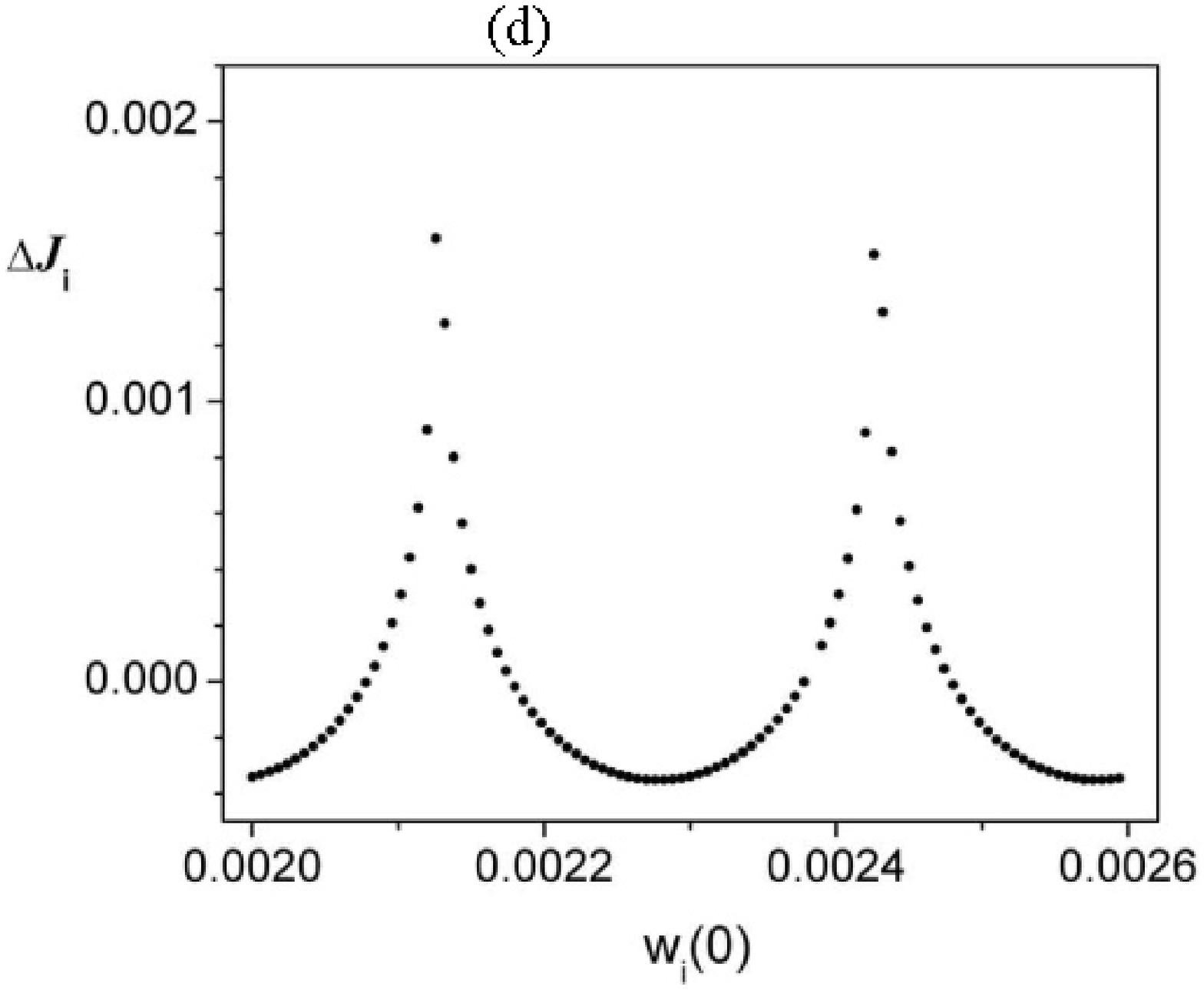} \caption{Scattering at the separatrix crossing. a) Bunch of
trajectories with various (but close) initial conditions
undergoing jump of the improved adiabatic invariant at separatrix
crossing. Trajectories are mixed due to the jumps. b) $\eps-$
dependence of magnitude of jump of the improved adiabatic
invariant. For every value of $\eps$, we calculated a bunch of 80
trajectories from $\delta=10$ to $\delta=0$. Initial values of $w$
were chosen to be equidistantly distributed in the interval
[$0.96, 0.96+1.5\eps$]. The theory predicts quasi-random jump of
the improved adiabatic invariant, which magnitude scales linearly
with $\eps$. We calculate mean value $\sigma$ of squared change in
the improved adiabatic invariant, which turns out to scale
perfectly linearly with $\eps$ (accordingly, dispersion $\sigma^2$
scales linearly with $\eps^2$) c) The same as in Fig.b, but with
smaller values of $\eps$ and initial values of action. We
calculated slope of the line $\sigma(\eps)$ taking into account
the four points with the smallest values of $\eps$, and get the
value $k \approx1.1614$, which is in good correspondence with
theoretical  prediction of $\sqrt{4/3} \approx 1.15$; for larger
values of $\eps$, correspondence worsens: $k \approx 1.015$ when
taking into account all points. d) High sensitivity of the jump of
the adiabatic invariant on initial conditions. Calculations for
$\eps=0.0004$ are presented. Initial values of $w$ for 100
trajectories were uniformly distributed in the tiny interval
($w_0$,$w_0+1.5\eps$). Change in the improved adiabatic invariant
was calculated ($\Delta J=J(\delta=0)-J(\delta=10)$). It is seen
that tiny change in the initial conditions results in large
variance of the jump of the action. Trajectories arrive at the
separatrix with different values of the pseudo-phase $\xi$. Maxima
in the Figure correspond to $\xi=0$ and $\xi=1$. The formula for
the jump of the adiabatic invariant predicts high increase in the
value of the jump when $\sim (\pi \xi)$ nears 0. In the very
vicinity of $\xi=0,1$ the formula is not working (the predicted
jump diverges while the calculated jump is finite), however
measure of the exceptional initial conditions leading to $\xi=0,1$
is very small \cite{AKN}. } \label{multiplej}
\end{figure*}

In the Case II the phase portraits have richer structure (Fig.
\ref{AMBECportrait}). With $\lambda \ne 0$, another saddle point
can appear at $\theta=\pi$ provided $\lambda < \lambda_c =
-\frac{1}{2^{3/2}}$. The appearance of this saddle point can be
understood from the graphical solution of the equation (see also
 \cite{Links}): \be 2\lambda w - \delta= -\frac{3 w +1}{2
\sqrt{w+1}}. \label{FP} \ee As $\delta$ is decreased, the line
$y(w)= 2\lambda w - \delta$ goes up and  crosses the curve
determined by the r.h.s of (\ref{FP}). Two points of intersection
represent the saddle point (which moves to $w=1$ as $\delta$ is
decreased further) and the elliptic fixed point which moves to
$w=-1$. As the saddle point reaches the $w=1$ segment, another
bifurcation occurs and the saddle point "splits" into the two
saddle points (similar to those in Fig.\ref{fermi}), that move
apart from $\theta=\pi$ along the segment $w=1$ and disappear at
$\theta=0$.

In Section IIb the dynamical change in the action in the case
$\lambda=0$ is considered in detail, while Section IIc briefly
discusses the case $\lambda \ne 0$ (geometric jump).

\subsection{Case I: negligible mean-field interactions,
$\lambda=0.$ Dynamical change in the action at the separatrix
crossing.}

Consider in a greater detail the passage through the separatrix in
Fig. \ref{fermi} described in the previous subsection. At large
positive $\dt$, $1-w$ is proportional to classical action, while
at large negative $\dt$ action is proportional to $1+w$ (see also
Fig. \ref{Bloch}). In the adiabatic limit, $w$ reverses its sign
due to passage through the resonance: the final and initial values
of $w$ are related as $w_{f}=-w_{in}$. Calculating change in the
action due to separatrix crossing (Refs. \cite{Comment,PhysD}),
one obtains the nonadiabatic correction to this adiabatic result.
It scales linearly with $\eps$ if initial population imbalance
slightly deviates from 1 (i.e., initial molecular fraction is not
very small).

 As the trajectory nears the separatrix due to slow change  (of
order $\eps$)in the parameter, the action undergoes oscillations
of order of $\eps$. Each oscillation corresponds to one period of
motion of the corresponding trajectory in the unperturbed system.
In the vicinity of separatrix, the period of motion grows
logarithmically with energy difference $h$ between energy level of
the unperturbed trajectory and the energy on the separatrix (so as
$h$ tends to 0, the period of motion tends to infinity). As a
result, the "slow" change of the parameter becomes "fast" as
compared to the period of motion: breakdown of adiabaticity
happens; oscillations of the adiabatic invariant grow and at the
crossing its value  undergoes a quasi-random jump (Fig.
\ref{singlejump}).

According to the general theory, it is not enough to consider
dynamics of the action variable. One introduces the improved
adiabatic invariant $J= I +\eps f(w,\theta,\tau)$  (see the
Appendix for brief description of adiabatic and improved adiabatic
approximations and the general formula for $J$ ). The improved
adiabatic invariant is conserved with better accuracy: far from
the separatrix, it undergoes very small oscillations of order
$\eps^2$. At the separatrix crossing, it undergoes jump of order
$\eps$. \newline We illustrate this behavior in Fig.
\ref{singlejump}. Figs. \ref{singlejump}a,b give dynamics of the
action (adiabatic invariant) $I$. It is clearly seen that before
and after separatrix crossing it oscillates around different mean
values, but the jump in action is of the same order as its
oscillations close to the separatrix.  Fig. \ref{singlejump}c
presents time evolution of the improved adiabatic invariant. The
jump in $J$ is much more pronounced (although it is possible to
express the improved adiabatic invariant in the elliptic
functions, we choose to calculate it numerically according to the
definition given in the Appendix A).

Now, at large $|\delta|$ not only the action $I$ coincides with
value of $1-|w|$, but also the improved adiabatic invariant $J$
coincides with $I$. Therefore, calculating change in the improved
adiabatic invariant $J$, we obtain change in the action and change
in the value of $1-|w|$ due to the resonance passage. For the case
of small initial action $I$, the change in action was calculated
in  \cite{Comment} according to the general method of \cite{AKN}.
The formula is \be 2\pi \Delta J = -2 \frac{\eps
\Theta_*}{\sqrt{2-\dt_*^2}} \ln(2\sin \pi \xi), \label{jumpp} \ee
where $\Theta$ is rate of change of the area within the separatrix
loop: $\Theta=\frac{dS}{d \tau}$ (note that the rate do not depend
on $\eps$); $\xi$ is the pseudo-phase: $\xi=|h_0/\eps \Theta|$,
where $h_0$ is the value of the energy at the last crossing the
vertex bisecting the angle between incoming and outgoing
separatrices of the saddle point $C$ outside the separatrix loop
(see Fig.\ref{fermi}c). Similar calculations were done  in
\cite{Chaos}. The formula can be further simplified by expressing
$\Theta$ via $\frac{\partial \delta}{\partial \tau}  \equiv
\delta'$. Indeed, the area within the separatrix loop is $
S(\delta) = 2 \int_{\delta^2 - 1}^1 dw \Bigl[ \pi- \arccos \left(
\frac{\delta}{\sqrt{1+w}}\right) \Bigr].$ We are interested in
derivative of $S(\delta)$ over $\delta$. Differentiating the above
integral over parameter $\delta$, one obtains: \be S'(\delta) =
4\sqrt{2-\delta^2}, \quad\mbox{and}\quad \Theta =
S'(\delta)\frac{\pt \delta}{\pt \tau} = S'(\delta) \delta'.   \ee
Therefore, the formula (\ref{jumpp}) is simplified to \be  \Delta
J = - \frac{4\eps \delta'}{\pi} \ln( 2\sin \pi \xi). \label{jump}
\ee

We carefully checked numerically behaviour of jumps in action
predicted by formula (\ref{jump}), see Fig.(\ref{multiplej}).
Fig.\ref{multiplej}a demonstrates scattering at the separatrix
crossing: bunch of trajectories with various (but close) initial
conditions undergoing jumps of the improved adiabatic invariant at
separatrix crossing. Figures  \ref{multiplej}b,c demonstrate
$\eps-$dependence of jumps of the improved adiabatic invariant.
For several values of $\eps$, a bunch of 80 trajectories with
close initial conditions was calculated, and dispersion of actions
due to separatrix crossing was calculated, which scales linearly
with $\eps^2$ (i.e., $\sigma^2 = K \eps^2$). Note that from the
formula (\ref{jump}) it is possible to determine not only the
linear power-law, but also the corresponding coefficient of
proportionality $K$. The theory predicts uniform distribution of
$\xi$, therefore dispersion of jump in the action can be
calculated as

\be \sigma^2 = 16 \eps^2 (\delta')^2 \pi^{-2} \int_0^1 \ln^2 2
\sin \pi \xi d \xi = \frac{4 \eps^2 (\delta')^2}{3} \ee

For numerical calculations, we used linear sweeping of $\delta$,
therefore the predicted dispersion is  $ \sigma^2 = 4\eps^2/3$.
Predicted coefficient $4/3$ can be compared with the slope in
Figs.\ref{multiplej}b,c . For relatively large $\eps$ (Fig.
\ref{multiplej}b), correspondence is not very good, but when we
decreased the value of $\eps$, we obtained $K  \approx 1.348$
which is in good correspondence with theoretically predicted $K =
4/3 \approx 1.333$. We reveal also high sensitivity of the jump of
the adiabatic invariant on initial conditions
(Fig.\ref{multiplej}d), which is the cause of uniform distribution
of $\xi$  \cite{AKN}. We therefore  checked almost all qualitative
and quantitative aspects of destruction of adiabatic invariance at
separatrix crossings in that model. Let us finally mention the
main steps in obtaining the formula \label{jump0}:
\begin{enumerate}
\item
Linearization around the saddle point in the frozen system and
derivation of a approximate formula for the period of motion $T$
along the trajectory with energy $h$. The period depends
logarithmically on $h$ and is inversely proportional to the square
root from the Hessian of the Hamiltonian in the saddle point
(determinant of the matrix of second derivatives).
\item
Obtaining the action variable $I$ from the period $T$ using the
formula $T= 2 \pi \partial I/\partial h$.
\item
Calculation the function $f$ at a point of the vertex bisecting
the angle between incoming and outgoing separatrices of the saddle
point (Fig. 1c). It is proportional to $\Theta$ (for details, see
 \cite{AKN}).
\item
"Slicing" the exact trajectory on parts (corresponding to "turns"
in the unperturbed system) by the bisecting vertex and
constructing a map $\tau_n, J_n \to \tau_{n+1}, J_{n+1}$ using the
analysis described above ($\tau_0$ is the moment of last crossing
of the vertex before the separatrix crossing, $\tau_{-1}$ is a
previous moment of crossing the vertex, etc. $J_n$ is value of the
improved adiabatic invariant at $\tau_n$). Summation of changes of
adiabatic invariant at each turn leads to the formula
(\ref{jump}).
 \end{enumerate} See Refs. \cite{Comment,PhysD} for further details.
\subsection{Case II: Condensates with interactions, $\lambda \ne 0$. Analog of nonzero adiabatic tunnelling.}
Let us briefly consider the model with $\lambda < \lambda_c=
-\frac{1}{2^{3/2}}$. Separatrix crossing happens via another
scenario here (according to the motion of the fixed points
described in Section IIa). We give only qualitative discussion of
a possible new phenomenon. We plot the phase portraits at
different $\delta$ and fixed $\lambda$ in Fig.
(\ref{AMBECportrait}). Now, as $\delta$ is decreased, three
domains $G_{1,2,3}$ appear in the phase portrait at certain
$\tau=\tau_*$ as a result of the first bifurcation. Shortly after
the bifurcation (see Fig. \ref{AMBECportrait}c) the separatrix
consists of the two "loops": the upper $G_2$ (adjacent to $w=1$
line), which area $S_2(\tau)$ decreases from $S_2(\tau_*)$ to zero
as the unstable fixed point goes towards $w=1$, and the bottom
$G_3$, whose area $S_3(\tau)$ increases from zero; $G_1$ is the
"outer" domain adjacent to $w=-1$.

In case initial action $I_0$ of a phase point is sufficiently
small $(2 \pi I_0 < S_2(\tau_*))$, the phase point resides in the
$G_2$ domain when the separatrix emerges (without any separatrix
crossing, see Fig.\ref{AMBECportrait}c). In case $2 \pi I_0$ is
larger then the area $S_2$ of the domain $G_2$ at the moment of
separatrix creation, the phase point occupy $G_1$ at this moment.
Consider the former case, i.e. small initial action. As $\delta$
evolves, $S_2$ decreases, while $S_3$ grows. When $S_2(t)$ becomes
equal $2 \pi I_0 $, separatrix crossing occurs and the phase point
is expelled to $G_1$ domain and then almost immediately to $G_3$
domain (say, in the Fig. \ref{AMBECportrait}f). It is easy to see
that the phase point acquires large action due to geometric jump
in the action when entering $G_3$, so in the end $w$ will deviate
from the all-molecule mode $w=-1$ considerably (the geometric jump
is equal to $(S_3(\tau_{**})- S_2(\tau_{**}))/2 \pi$, where
$\tau_{**}$ is the moment of the separatrix crossing). This is in
some sense analogous to the {\em nonzero AT} discussed in
\cite{Niu,Zobay} and in Section III of the present paper. One
might try to explain the sizable remnant fraction after the
adiabatic Feshbach resonance passage as the geometric jump in the
action due to the self-trapping effect of s-wave interactions.
This, however, requires further investigation: while calculation
of the geometric jump in action is trivial, dynamical jump is not
so easy to calculate in this geometry. So far, we just suggest a
possible new phenomenon in the model, detailed discussion will be
given elsewhere.

\begin{figure*}
\includegraphics[width=160mm]
{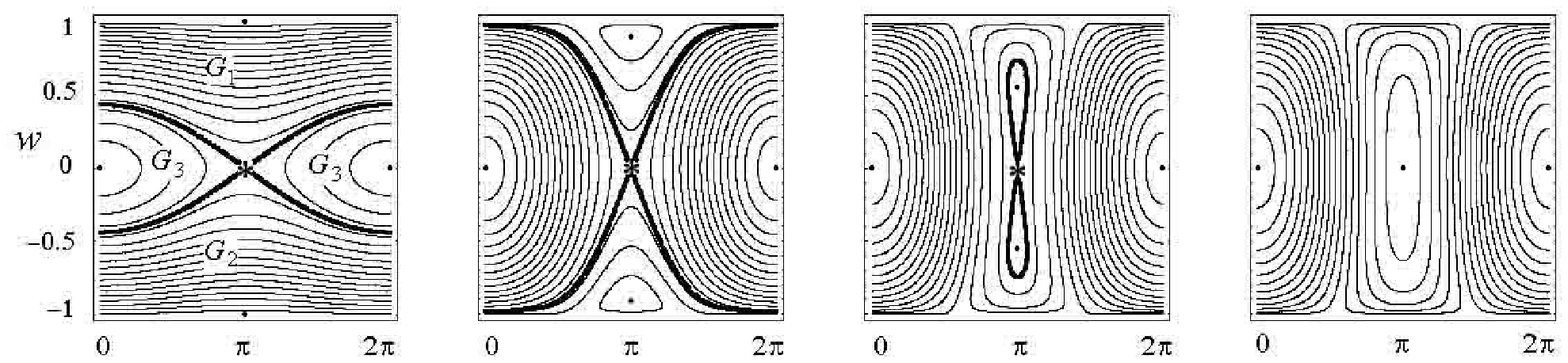} \caption{Phase portraits of the 2-mode ABEC Hamiltonian with
$\delta=0$. From left to right (a,b,c,d):
$\lambda=20,2.4,1.2,0.8$. As $\lambda$ decreases, separatrix loop
grows until $\lambda=2$ where it changes its configuration, and at
$\lambda=1$ it disappears. On the other hand, by increasing
$\lambda$ it is possible to switch from regime of complete
oscillations (domain 3) to the self-trapped regime (domains 1 or
2). The unstable fixed point do not move: it is either at (0,0) or
absent. \label{ABECsymm} }
\end{figure*}

\begin{figure}
\includegraphics[width=60mm]
{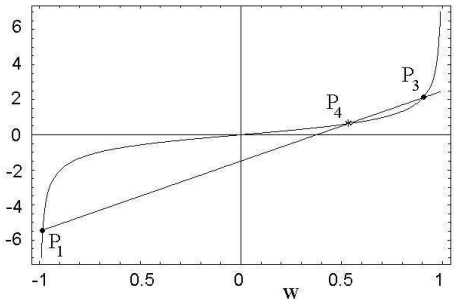} \caption{Graphical solution of the equation $-\delta+\lambda
w = w/\sqrt{1-w^2}$ which gives fixed points at $\theta=\pi$. As
$\delta$ decreases, the line goes up, and three fixed points can
appear from a single one at certain window of value of $\delta$
provided $\lambda>1$. The star denotes the unstable fixed point
which after the birth goes down and collides with the stable fixed
point. See corresponding phase portraits in the next Figure.
\label{FNLZg} }
\end{figure}

\begin{figure*}
\includegraphics[width=140mm]
{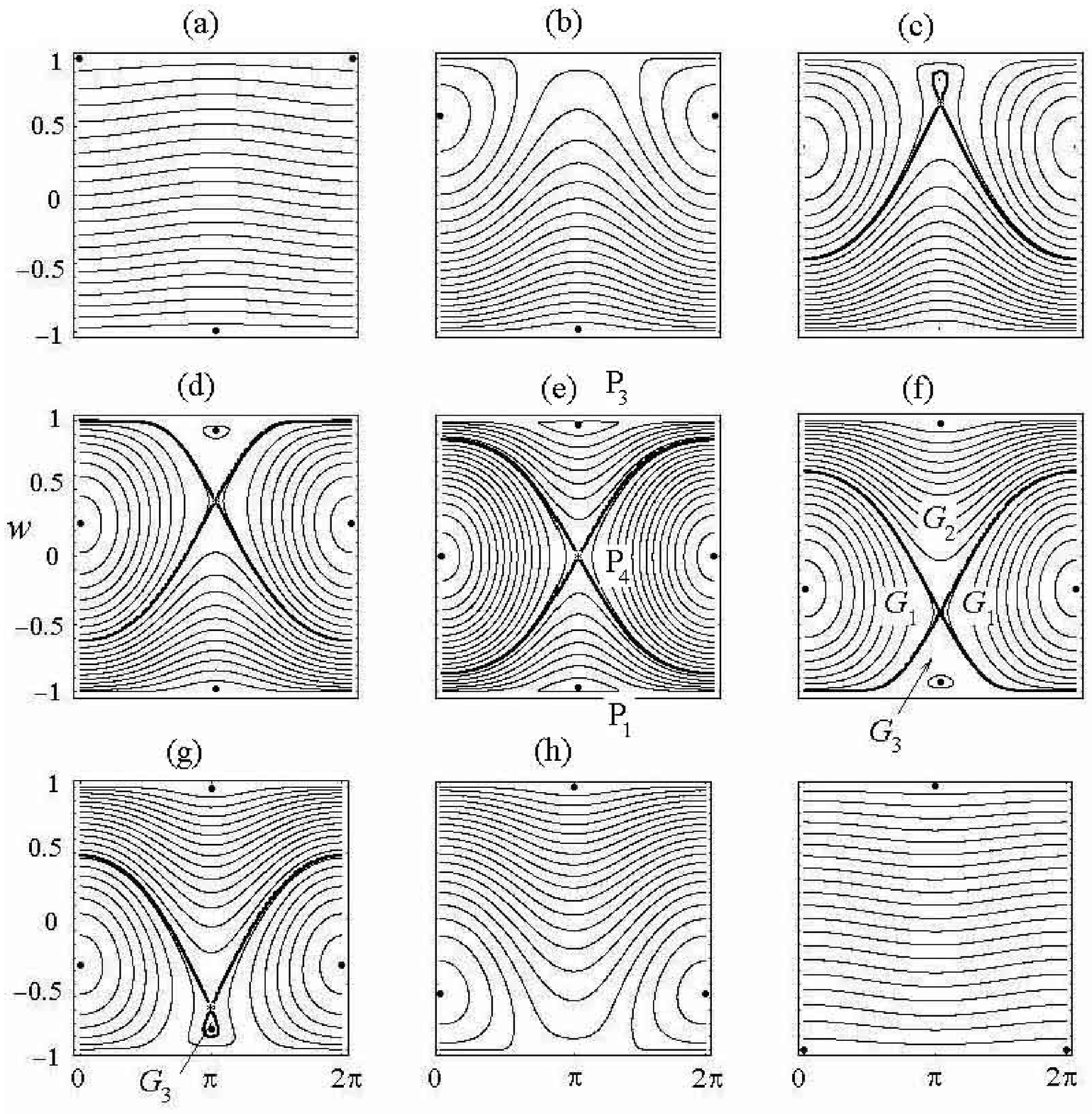} \caption{Nonlinear Landau-Zener tunneling: phase portraits
of the 2-mode ABEC Hamiltonian at different values of $\delta$.
From top left to bottom right:
$\delta=20,3,1.8,1.2,0,-1.2,-1.8,-3,-20$; $\lambda$=const=4.
\label{FNLZ}}
\end{figure*}

\section{Nonlinear two-mode model for two coupled BEC.}
\subsection{Model equations and its physical origin; phase portraits}

We consider the Hamiltonian ("nonlinear 2-mode ABEC model")\be
H=-\delta w+ \frac{\lambda w^2}{2}-\sqrt{1-w^2}\cos\theta  \label
{2ABEC}\ee

There are many systems in BEC physics that are described in the
mean-field limit by the Hamiltonian (\ref{2ABEC}). It has been
used to model two coupled BECs (BEC in a symmetric double well in
case $\delta=0$) (\cite{Smerzi}). The model with $\delta \ne 0 $
is equivalent to nonlinear Landau-Zener model, which appears, in
particular, in studying BEC acceleration in optical lattices
\cite{Zobay,Niu}

Theory of nonlinear Landau-Zener tunnelling was suggested in
\cite{Niu,Zobay}. However, only the case of zero initial action
was considered. In particular, it is said in \cite{Niu} that
adiabaticity is broken when "fixed points collide". In case the
initial action is not zero, adiabaticity is broken before that: it
is broken when separatrix crossing occurs. Therefore, it is
necessary to involve theory of separatrix crossings in
consideration of these models.

It is worth to mention that for BEC in a symmetric double-well,
there exist also improved 2-mode model \cite{Bergman}, where the
term $\cos 2 \phi$ is added: \be H= A \frac{z^2}{2} - B
\sqrt{1-z^2} \cos \phi + \frac{1}{2}C (1-z^2) \cos 2 \phi, \ee
where parameters $A,B,C$ are determined by overlap integrals of
the mode functions. Usually, the $\cos 2 \phi$ term is small and
can be omitted. Then, the improved model Hamiltonian can be
reduced to (\ref{2ABEC}) with $\delta=0$ (still, coefficients are
determined more accurately in the improved model). The original
model is derived for the case of constant parameters. One may
wonder if it is working in a time-dependent situation. It is not
difficult to demonstrate that for slowly changing parameters one
can use the same model, with parameters of the Hamiltonian slowly
changing in accordance with the "instantaneous" model. For
simplicity, let us demonstrate this using the improved 2-mode
model \cite{Bergman} as an example. The order parameter in a
two-mode approximation is \bea \psi(x,t)= \sqrt{N}[\psi_1(t)
\Phi_1(x) + \psi_2(t) \Phi_2(x)], \label{expansion}\\
\Phi_{1,2}(x)= \frac{\Phi_+(x) \pm \Phi_(x)}{\sqrt{2}},
\nonumber\eea where $\Phi_{\pm}$ satisfy the stationary GP
equation \be \beta_{\pm} \Phi_{\pm} = -\frac{1}{2} \frac{d^2
\Phi_{\pm}}{dx^2}+ V_{\mbox{ext}} \Phi_{\pm} + g |\Phi_{\pm}|^2
\Phi_{\pm} \label{gp}\ee

The variables of the classical Hamiltonian are defined as \be
z(t)= |\psi_1(t)|^2- |\psi_2(t)|^2, \quad
\phi(t)=\mbox{arg}\psi_2(t)-\mbox{arg}\psi_1(t)  \nonumber
 \ee
Substituting (\ref{expansion}),(\ref{gp}) into the time-dependent
GP equation, one gets \cite{Bergman} \bea i \frac{d \psi_1(t)}{dt}
(\Phi_+ &+& \Phi_-) + i \frac{d
\psi_2 (t)}{dt} (\Phi_+ - \Phi_-) = \nonumber\\
\Sigma_{\pm}(\psi_1(t) \pm \psi_2(t))[ \beta_{\pm} &-& gN
|\Phi_{\pm}|^2 ] \Phi_{\pm} \ +\\
\frac{gN}{2} \Sigma_{\pm}[\Phi_{\pm}^3 P_{\pm} &+& \Phi_{\pm}^2
\Phi_{\mp} Q_{\pm} ], \nonumber \eea where $P_{\pm},Q_{\pm}$ are
functions of $\psi_1,\psi_2$ (see \cite{Bergman}). From these
equations, one get the equations of motion for $\psi_1,\psi_2$
(Eqs. 13 from \cite{Bergman}):

\bea i \dot{\psi_{\mu}} =(F+A |\psi_{\mu}|^2 - \frac{\Delta
\gamma}{4} \psi_{\mu} \psi^*_{\nu} ) \psi_{\mu} +
\nonumber\\(-\frac{\Delta \beta}{2} + \frac{\delta \gamma}{4}
|\psi_{\mu}|^2 + C \psi^*_{\mu} \psi_{\nu} ) \psi_{\nu}, \eea
which can be rewritten as Hamiltonian equation of motion of the
corresponding classical pendulum ($F,A,C,\Delta \gamma, \Delta
\beta$ are functions of mode overlap integrals and energies
$\beta_{\pm}$). Considering time-varying parameters, we introduce
instantaneous mode functions $\Phi_{\pm}(x,t)$. If we keep
two-mode expansion of the order parameter, when it is not
difficult to show that additional terms coming from
time-dependence of the mode functions ($\int \Phi_+ \frac{\partial
\Phi_+ }{\partial t}d\mathbf{r} $, $\int \Phi_+ \frac{\partial
\Phi_- }{\partial t}d\mathbf{r}  $, etc ) are strictly zero due to
symmetry and normalization conditions. Complications can arise
only from excitation of other modes (if we would allow, say,
four-mode expansion). However, we do not consider this question
here. Even in the two-mode approximation nonadiabatic dynamics is
nontrivial, and it comes purely from nonadiabatic behaviour of
classical action. Phase portraits of the model (\ref{2ABEC}) with
$\delta=0$ are given in Fig. \ref{ABECsymm}. We are interested
only in the supercritical case here. Separatrix crossings and
corresponding changes in the action are discussed in Section IIb.
The case $\delta \ne 0$ (NLZ model) is discussed in Section IIc,
where we present a new phenomenon: {\em separated AT}.

\begin{figure*}
\includegraphics[width=100mm]
{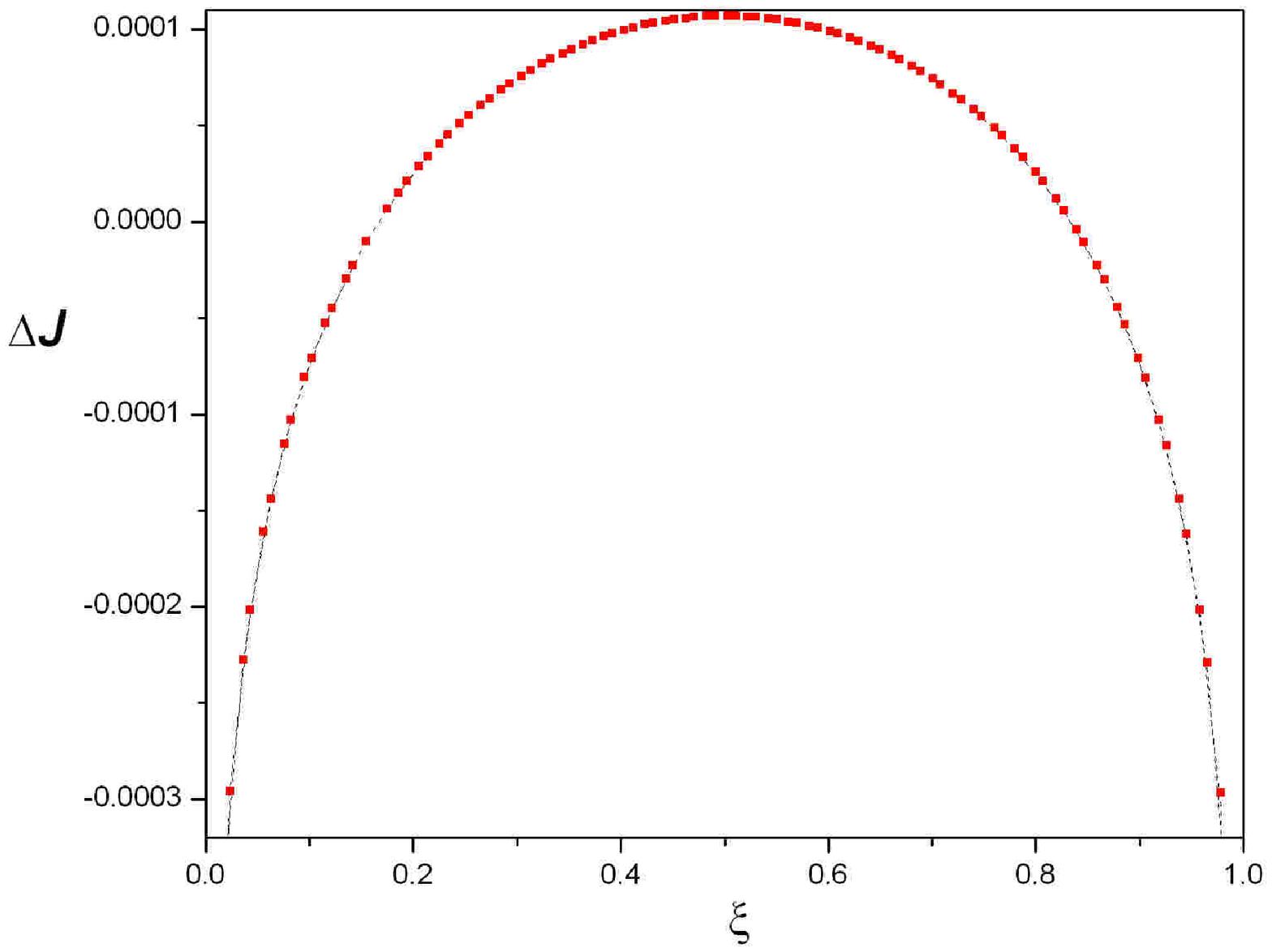}
\includegraphics[width=100mm]
{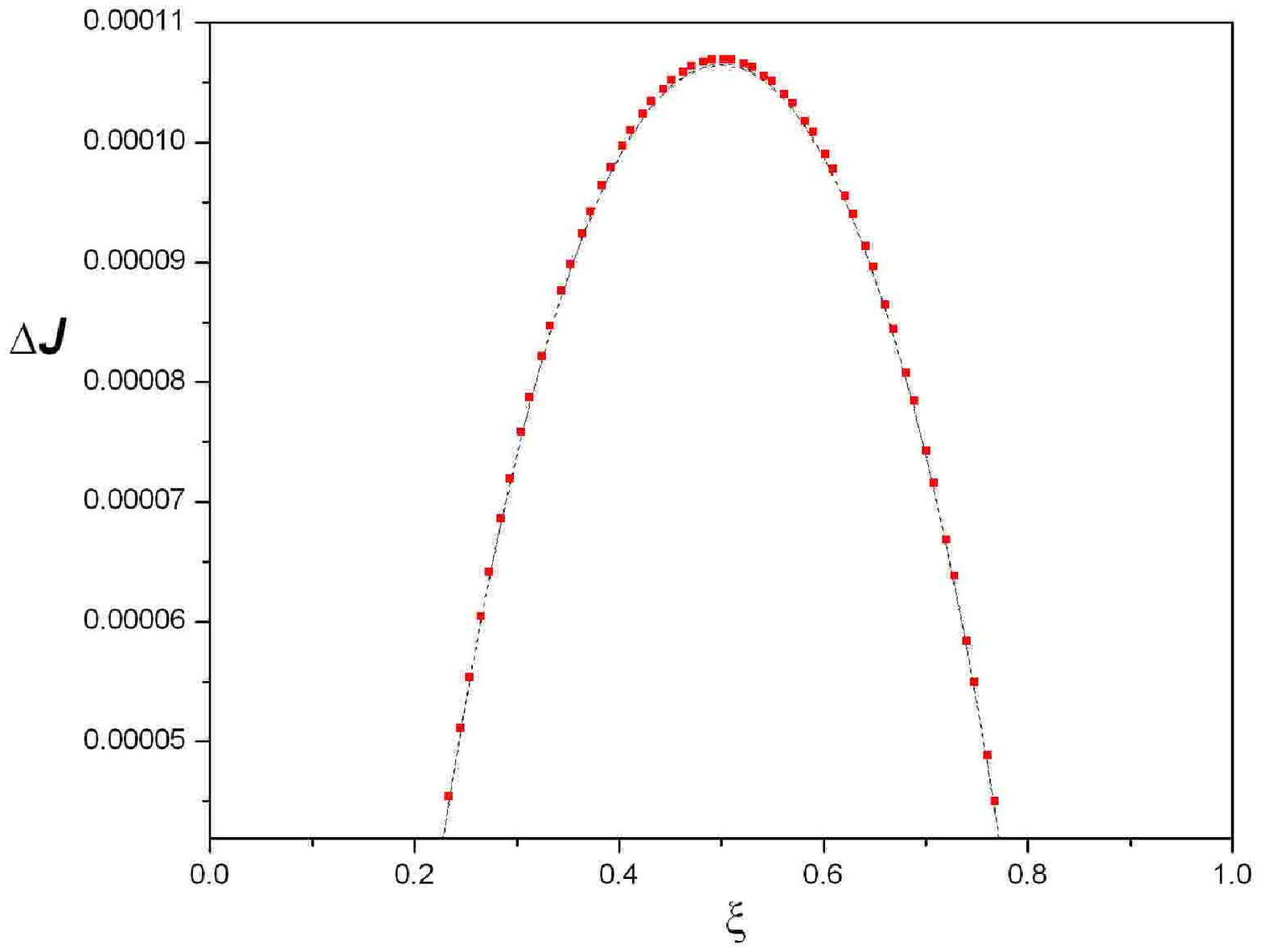}
 \caption{Jump in the improved adiabatic invariant in dependence of the pseudo-phase $\xi$. Filled squares: numerical results; dashed line:
 analytical predictions according to Eq. \ref{symmjump}. We slowly changed $\lambda$ according to the law $\lambda=\lambda_a-\lambda_b \cos \eps t$, with $\eps=0.001$,
  $\lambda_a=15$,$\lambda_b=10$. We took a set of 100 phase points with different but very close initial conditions: $w_i=0$, $\theta_i$ are distributed along an interval
  of order of $\eps$ at the time $\tau=\eps t$=0. We propagate the bunch of trajectories until the time $\tau=\pi$ (so all the points changed
  its regime of motion from complete oscillations to the self-trapped mode). For each
  point, value of $\xi$ and change in the improved adiabatic invariant $\Delta J$ was determined numerically, then
  the analytical prediction for the change in the improved adiabatic invariant $\Delta J(\xi)$  was calculated according to Eq.\ref{symmjump}.
  The numerical and analytical results shown in the Fig. (a) are almost indiscernible, in (b) enlarged part of the same plot is presented,
  where small deviations are seen. It is also important to mention, that from 100 phase points exactly 50 were
   trapped in the upper domain $G_1$, and 50 in the lower $G_2$.  \label{symmjumpcheck}}
\end{figure*}

\subsection{Case I: symmetric double-well, $\delta=0$.}

We suppose initially the system is in the oscillating regime of
complete tunnelling osciallations (domain $G_3$), and then due to
slow change of parameters is switched into self-trapped regime.
Two different probabilistic phenomena take place at the crossing:
quasi-random jump in the action and the probabilistic capture.

Consider the probabilistic capture: there are two domains
$G_{1,2}$ for the self-trapped regime in the phase portraits: in
the first (upper) $w>0$, in the second (bottom) $w<0$. In which of
these two domains the phase point will be trapped (in other words,
in the left or the right well)? The trapping in either of the
domains is also very sensitive to initial conditions; in the limit
of small $\eps$ the trapping is a probabilistic event. For the
symmetric case, the probability to be trapped in ether well is
exactly $1/2$. However, for the asymmetric well the answer is not
so straightforward. It is determined by some integrals over
separatrix at the moment of switching (general theory exists, see  \cite{AKN}). %

As for the first phenomenon (jump in the action), at the moment of
switching, destruction of adiabaticity happens in the sense that
the adiabatic invariant undergoes a relatively large jump of order
of $\sqrt{\eps}$ (very similar to that discussed in the Section
II). If we then slowly bring the parameters back to the initial
values, the adiabatic invariant will be different.

The formulas for the action-angle variables are cumbersome. In
fact, to calculate change in the action, it is not necessary to
have formulas for the action-angle variables. The jump is
determined by local properties of the Hamiltonian near the
separatrix: the area of the separatrix loop and the Hessian of the
unstable fixed point \cite{AKN}. As a result, the formula for the
jump of the action is simplier than expressions for the action
itself. Suppose $\lambda
> 2$ so the phase portrait looks like in Fig. \ref{ABECsymm}b and
we start from the regime of complete oscillations. Slowly changing
$\lambda$, we can switch to the  self-trapped regime. The
expression for the area of the separatrix loop is easy to
calculate: \be S(\tau)/4=b+ \arcsin b, \quad
b=\frac{2\sqrt{\lambda-1}}{\lambda} \ee

The Hessian of the Hamiltonian in the unstable fixed point can be
calculated as $ D(\tau)=-(\lambda-1)$.

Let us define \be
 d(\tau) \equiv  1/\sqrt{-D(\tau)},
\ee

We calculated jump of the action according to the general method
as

\be \Delta J = -\frac{1}{2 \pi} \eps d_* \Theta_* \ln(2 \sin (\pi
\xi ))= \eps \frac{4 \lambda'}{\pi \lambda^2} \ln(2 \sin (\pi \xi
)) , \label{symmjump} \ee

where $\xi$ is the pseudophase corresponding to the first crossing
of line $\theta=\pi$ in the $G_{1,2}$ domains, $d_*$ is value of
$d$ at the moment of crossing this line, values of $\lambda$ and $
\lambda'$ are also taken at this moment.

We checked this formula numerically. A set of 100 phase points
with initial conditions being distributed in a small (of order
$\eps$) interval far from the separatrix were chosen. Then, the
bunch of trajectories in the system with slowly changing parameter
was calculated. For each trajectory, values of $\xi$ and $\Delta
J$ (change in the improved adiabatic invariant) were determined.
From numerically determined $\xi$, theoretical prediction for
change in the action $\Delta J$ was calculated and compared with
numerically determined $\Delta J$. Results are in the Fig.
\ref{symmjumpcheck}; correspondence between numerical results and
analytical prediction is perfect. In the same calculations,
mechanism of quasi-random division of phase flow was verified:
exactly one half of the phase points from the considered set were
captured in the upper domain $G_1$, and the other half were
trapped in the lower domain $G_2$. This is a purely classical
phenomenon, the sound example of probabilistic phenomena in
dynamical systems ( \cite{AKN,Arnold63}).

\subsection{Case II: asymmetric double-well and nonlinear Landau-Zener model, $\delta \ne 0$.
{\em Separated adiabatic tunnelling}.}

Consider sweeping value of $\delta$ from large positive to large
negative values in Fig.\ref{FNLZ} (see also  \cite{Zobay,Niu}). In
case $\lambda<1$, only two fixed points exist at $\theta=0,\pi$
($P_2,P_1$ correspondingly). As $\delta$ changes from
$\delta=-\infty$ to $\delta=+\infty$, $P_1$ (corresponding to the
lower "eigenstate") moves along the line $\theta=\pi$ from the
bottom $(w=-1)$ to the top $(w=1)$, the other point $P_2$
(corresponding to the upper "eigenstate") moves from the top to
the bottom. In case $\lambda>1$, two more fixed points appear in
the window $-\delta_c<\delta<\delta_c, \quad
\delta_c=(\lambda^{2/3}-1)^{3/2}$. We concentrate on this,
"above-critical" case. The new points lie on the line
$\theta=\pi$, one being elliptic ($P_3$) and the other hyperbolic
($P_4$). Again, it is convenient to use graphical solution (Fig.
\ref{FNLZg}) to visualize appearance and disappearance of the
fixed points. It is stated in the \cite{Niu}, that collision
between $P_1$ and $P_3$ leads to nonzero AT from the lower level
to the upper level, and tunnelling probability in the adiabatic
limit is obtained by calculating phase space area below the
"homoclinic trajectory" ( which is the limiting case of the
separatrix with $S_3$=0), i.e. as {\em geometric} jump in the
action. In the zeroth order approximation, this approach is
correct (if the initial action is zero or very small). It is very
important that we can adopt general theory of separatrix crossings
to the case of this model with nonzero initial action
(corresponding to initially excited system). In this case
destruction of adiabaticity happens {\em before} collision of the
phase points. Initial trajectory is almost a straight line, so the
initial action is equal to $w+1$ in case we start close to $w=-1$,
or $1-w$ in case we start close to $w=1$. Consider the former
case.  Let the initial action $I_0$ (i.e., value of $w+1$ in Fig.
\ref{FNLZ}a) be equal to area of the separatrix loop in Fig.
\ref{FNLZ}g. The phase point is oscillating around slowly moving
$P_1$ point until the area of the separatrix loop $S_1(\tau)$
becomes equal to $2 \pi I_0$ at some moment $\tau=\tau_*$. Where,
separatrix crossing occurs. Action undergoes geometric jump (which
is simply  $[S_3(\tau_*)-S_1(\tau_*)]/ 2 \pi$. This geometric jump
is the analog of AT probability discussed in  \cite{Niu,Zobay} for
the case of zero initial action. The geometric jump is accompanied
by the {\em dynamical} jump similar to that discussed in Section
II and Section IIIb. The dynamical jump is small (of order of
$\eps$) as compared to the geometric jump. But conceptually it is
very important: only dynamical jump leads to destruction of
adiabatic invariance in the model. Indeed, if we reverse change in
$\delta$, the phase point will return to its initial domain and
the geometric jump will be completely cancelled. However,
dynamical jumps will not be cancelled, and at multiple separatrix
crossings they lead to slow chaotization (see, for example,
\cite{Bil}). Formulas for the dynamical jumps in the asymmetric
case are more complicated, as there are terms of order $\eps$ and
$\eps$ $\mbox{ln} \eps$.
 However, the probabilistic capture in this case is very much different. Consider the phase
portraits in Figs. \ref{FNLZ}f,g. Suppose that not only $\delta$,
but also $\lambda$ is changing. At the moment of crossing, the
area $S_3$ is diminishing, while the areas $S_{1,2}$ can behave
differently depending on evolution of parameters. Suppose both
$S_{1,2}$ are increasing: $\Theta_{1,2}>0, \quad \Theta_3<0 $.
Denote as $l_{1,2}$ the parts of the separatrix below and above
the saddle point, correspondingly. There is phase flow across
$l_2$ from the domain $G_2$ to $G_1$, and across $l_1$ from $G_3$
to $G_2$. The latter flow is divided quasi-randomly between $G_2$
and $G_1$: the phase point leaving $G_3$ can remain in $G_2$ or be
expelled to $G_2$. This is "determined" during the first turn
around the separatrix. After that, the particles are trapped
either in $G_1$ or $G_2$. Probability for either event can be
calculated as integrals over the separatrix parts $l_{1,2}$ (
\cite{AKN}): \bea {\cal} {\cal P}_1&=&\frac{I_2-I_1}{I_1}, \qquad
{\cal
P}_2=\frac{I_2}{I_1}, \label{probability}\\
 I_i (\delta,\lambda) &=& \oint_{l_i} dt \frac{\partial
\bar H}{\partial \rho}=\oint_{l_i} dt \left( \frac{\partial
H}{\partial \rho} -\frac{\partial  H_s}{\partial \rho} \right),
\quad \rho=\eps t. \nonumber \eea Here integrals are taken along
the unperturbed trajectories at the moment of separatrix crossing
(or last crossing the line $\theta=\pi$ before the separatrix
crossing), $H_s$ is the (time-dependent) value of the Hamiltonian
$H$ in the unstable fixed point, $\bar H$ denote the Hamiltonian
$H$ normalized in such a way as to make value of the new
Hamiltonian in the unstable fixed point to be zero. It is possible
to calculate all the integrals analytically, see the Appendix B.
We present numerical example in Fig. \ref{fcapture}. A set of
$N=100$ trajectories was considered with initial conditions
distributed in a tiny interval of $w$, and with $\theta(0)=0$ (so
initial actions were distributed in a tiny interval of order
$\eps$: $I_k=I_0+k \delta I$, $N \delta I \sim \eps, \quad
k=1,..,N$; alternatively, one can consider a set of phase point
with equal initial actions, but with distribution of phase along
$2 \pi$ interval). Both $\delta$ and $\lambda$ were changed; so
after the separatrix crossing a phase point can be trapped either
in $G_1$ or $G_2$. From the set of 100 points, 87 were trapped in
$G_1$, while 13 were trapped in $G_2$. The difference between the
final actions of these two subsets is approximately $I_0$, the
initial action of points in the bunch. The probability of 87\% is
in good correspondence with the theoretical prediction, which
gives $P_2=86,998$ for the probability of capture into the domain
$G_2$. Possible experimental realization of this new phenomenon is
again BEC acceleration in optical lattices, but with simultaneous
modulation of the lattice potential depth.

\begin{figure*}
\includegraphics[width=80mm]
{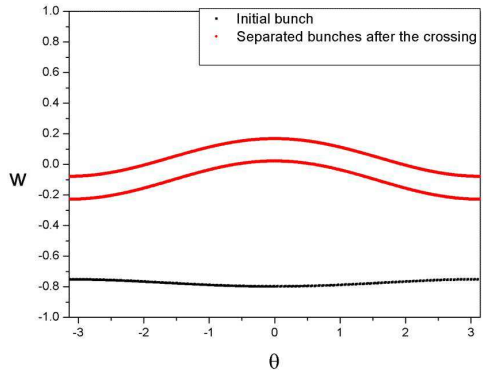}
 \caption{(Color online)Separated AT. We took a bunch of 100 trajectories with
 initial actions distributed in a tiny interval of order $\eps$ (in the Figure: the bottom curve consists of 100 initial trajectories;
  the two upper curves consist of 87 and 13 trajectories, correspondingly, and represent "snapshot" of the trajectories after
   the end of sweeping parameters).
 As parameters are changing, the separatrix (see  Figs. 9d-f) moves
 down, and crosses the bunch.  Due to quasi-random division of phase flow described in the text,
 some of the points were captured to the $G_1$ domain, while the other
 to the $G_2$ domain. As a result, phase points undergo different geometric change in the action. After the capture, actions
 are conserved. Therefore, a phase point can acquire two different
 values of the adiabatic invariant. The difference between the
 values corresponds to the area of the domain $G_3$ at the moment
 of separatrix crossing, i.e. it is approximately equal to the
 initial action. The question is how the initial bunch is divided, what is
 the probability for a phase point to come into either of the two upper bunches. From the set of 100 points, 87 were trapped in
 the upper bunch, while 13 in the bottom. This numerical result is in very good accordance
 with the theoretical prediction for the probabilities (\ref{probability},\ref{explicit}), which gives ${\cal P}_2=86.998$ (see the Appendix B).
 In case of nonzero {\em AT } considered in
 \cite{Niu,Zobay}, initial bunch would lie near $w=-1$ line, and
 there would be only one "final" bunch. Here, there are separated
 bunches, which suggested us to introduce the terminology "separated AT".
  \label{fcapture}}
\end{figure*}

\section{Conclusion}

We discussed destruction of adiabatic invariance in several
nonlinear models related to BEC physics. We especially
concentrated on the cases that have not been considered in the
corresponding papers on BEC dynamics yet: that is, when the
initial action is not zero.

We found that the general theory of adiabatic separatrix crossings
works very well in the considered models. Two aspects of
destruction of adiabatic invariance were considered: quasi-random
jumps in the approximate adiabatic invariants and quasi-random
captures in different domains of motion at separatrix crossings.

We discussed quasi-random jumps in the approximate adiabatic
invariants in the models describing Feshbach resonance passage in
coupled atom-molecule BECs, BEC tunnelling oscillations in a
double well, and nonlinear Landau-Zener tunnelling. Comparing with
previous analysis of the abovementioned models \cite{Vardi,Niu},
the key feature of our approach should be emphasized: the system
is linearized near the hyperbolic fixed point, not near elliptic
fixed points of the unperturbed system.

Another important class of phenomena considered here is
probabilistic captures into different domains of motion. They were
discussed for the case of BEC tunnelling oscillations in a
(symmetric or asymmetric) double-well and the NLZ model with
time-dependence of the nonlinearity $\lambda$. {\em Separated AT}
was discovered in the latter case. We suppose it can have
experimental applications in BEC manipulations with optical
lattices. The conceptual phenomenon of probabilistic capture was
firstly discovered in celestial mechanics (while studying
resonance phenomena in Solar system). It is interesting to draw an
analogy between intricate phenomena of celestial dynamics and
phenomena happening in many-body quantum systems. Conceptual
phenomena related to the classical adiabatic theory (which
includes both adiabatic invariants and the adiabatic (geometric)
phases) have recently become  important trend of research in the
highly interdisciplinary BEC physics field (see
\cite{PhysD,concept1,concept2,concept3}). We hope the
comprehensive analysis presented in this paper adds important
contribution to understanding nonlinear dynamics of Bose-Einstein
condensates.

\section{Acknowledgements}
A.P.I. was supported by JSPS.
 A.P.I. thanks D.Feder and A.Perali
for their help during participation in BCAM05 meeting and NQS-2005
conference correspondingly, where part of this work was presented.
Encouraging short discussions with R.Hulet, M.Oberhaler,
E.Arimondo, and W.Ketterle are acknowledged. Many thanks to A.A.
Vasiliev and A.I. Neishtadt for clarifying discussions and help in
the calculations. The authors thank O.I. Tolstikhin, S.Dmitriev,
Y.Kivshar and K.Nakagawa for discussions. This work was supported
also by Grants-in-Aid for Scientific Research No. 15540381 and
16-04315 from the Ministry of Education, Culture, Sports, Science
and Technology, Japan, and also in part by the 21st Century COE
program on ``Coherent Optical Science''.

\section{Appendix}
\subsection{Adiabatic and improved adiabatic approximations}

 To consider change in the action during a separatrix
crossing, it is necessary to introduce improved adiabatic
invariant $J$ in addition to the ordinary action variable $I$.
Improved adiabatic approximation is discussed in \cite{AKN}.

Let $I=I(w,\theta,\tau), \quad \phi=\phi(w,\theta,\tau)$ mod $2
\pi$ be the action-angle variables of the unperturbed
($\tau$=const) problem. The "action" $I(w,\theta,\tau)$ multiplied
by $2 \pi$ is the area inside the unperturbed trajectory, passing
through the point $(w,\theta)$ (provided the trajectory is closed;
otherwise the area of a domain bounded by the trajectory and lines
$\theta=0,2\pi$ is calculated). The "angle" $\phi$ is a coordinate
on the same unperturbed trajectory. It is measured from some curve
transversal to the unperturbed trajectories. The change
$(w,\theta) \to (I,\phi)$ is canonical (and can be done using a
generating function which depends on $\tau$). In the exact system
(with $\dot{\tau}=\eps \ne 0$) the variables $I$ and $\phi$ are
canonically conjugated variables of the Hamiltonian \be
H=H_0(I,\tau) + \eps H_1(I,\phi,\tau), \label{hh1}
 \ee
where $H_0(I,\tau)$ is the initial Hamiltonian $E(w,\theta,\tau)$
expressed in new variables, while the perturbation $H_1$ comes
from the time derivative of the generating function. In case the
angle $\phi$ is measured from some straight line $\phi=$ const,
one has the formula \cite{AKN} \be H_1= \frac{1}{\omega_0}
\int^{\phi}_0 \left( \frac{\partial E}{\partial \tau} -
\left<\frac{\partial E }{\partial \tau} \right> \right) d \phi,
\quad \omega_0=\frac{\partial H_0}{\partial I}, \ee where the
brackets $<..>$ denote averaging over the "angle" $\phi$.

Consider a phase point of the exact system with the initial
conditions $I=I_0, \phi=\phi_0$. The adiabatic approximation is
obtained by omitting the last term in (\ref{hh1}) and gives  \be
I=I_0, \quad \phi=\phi_0+ \frac{1}{\eps} \int_0^{\eps t}
\omega_0(I,\tau)d\tau \ee. Improved adiabatic approximation is
introduced in the following way. One makes another canonical
change of variables $(I,\phi) \to (J,\psi)$. The change is
$O(\eps)-$ close to the identity and in the new variables the
Hamiltonian has the form \bea H=H_0(J,\tau)+ \eps \bar H_1
(J,\tau)+ \eps^2 H_2(J,\omega,\tau,\eps),  \\ \bar H_1 = <H_1>=
-\frac{1}{\omega_0} \int_0^{2\pi} \left( \frac{1}{2} -
\frac{\phi}{2\pi}\right) \frac{\partial E}{\partial \tau} d\phi .
\label{hh2}\eea

The improved action variable can be defined as

\bea J &=& J(w,\theta,\tau)+I+\eps u,  \\
   u &=& u(w,\theta,\tau)=\frac{1}{2\pi} \int_0^T \left( \frac{T}{2}-t
   \right)\frac{\partial E}{\partial \tau} dt,
\eea where the integral is taken along the unperturbed trajectory
passing the point $(w,\theta)$, $T=\frac{2\pi}{\omega_0}$ is the
period of the trajectory, and the time $t$ is measured starting
from the point $(w,\theta)$. Determined in this way, $<u>=0$. The
improved adiabatic approximation is obtained by omitting the last
term in (\ref{hh2}) and gives

\bea J=J_0,  \quad \psi=\psi_0 &+& \frac{1}{\eps} \int_0^{\eps t}
\left( \omega_0(J,\tau) + \eps \omega_1 (J,\tau) \right)d\tau, \nonumber\\
\omega_1&=&\frac{\partial \bar F_1}{\partial J}. \eea

\subsection{Probabilities of captures during separated AT}

We change both $\delta$ and $\lambda$ linearly in time:
$\delta=\delta_0-\eps t$, $\lambda=\lambda_0-\kappa \eps t$,
$\kappa=1.5$; $\lambda_0=25$, $\delta_0=8$. We consider a bunch of
$N=100$ trajectories with initial conditions $w_k=w_0+ 0.02 \eps
k,$ $\theta_k=0$ ($w_0=-0.8$) which imply distribution of initial
actions in a tiny interval of order $\eps$. Alternatively, one can
consider initial conditions with the same initial action, but with
distribution along the angle variable $\phi$. In any case, from
$N$ trajectories, approximately ${\cal P}_2 N$ will be captured in
domain $G_2$, and ${\cal P}_1 N$ in domain $G_1$. As a result,
after sweeping value of $\delta$ to $-\infty$, one obtains two
bunches of trajectories each closely distributed along two
different values of action. This is a new phenomenon in the
context of nonlinear Landau-Zener tunnelling.

At the moment of separatrix crossing, phase portrait looks like
shown in Fig. \ref{FNLZ}f.  Phase flow from the domain $G_3$ is
divided between $G_1$ and $G_2$. It is possible to calculate
analytically the probabilities of captures in either domain.  The
separatrix crosses the line $\theta=0$ at points $w=w_{a,b}$,
$w_a<w_b$ and the line $\theta=\pi$ at $w=w_s$ (the unstable fixed
point). These three magnitudes ($w_{a,b,s}$) are the roots of the
equation

\be (\dot{w})^2=1-w^2-(h_s+\delta_* w - \frac{\lambda_*}{2}
w^2)^2=0, \label{3roots} \ee where $h_s$ is the energy on the
separatrix at the moment of crossing, and $\delta_*,\lambda_*$ are
values of the parameters at this moment ($w=w_s$ is the doubly
degenerate root). In other words,

\be \dot{w}= \pm \sqrt{-\frac{\lambda_*^2}{4}
(w-w_a)(w-w_b)(w-w_s)^2 } \label{3factor} \ee

Probabilities of capture in either domain are given by

\bea
{\cal P}_2&=&\frac{I_2}{I_1}, \quad {\cal P}_1=\frac{I_2-I_1}{I_1}, \nonumber\\
I_{1,2}= \frac{1}{2} \oint_{l_{1,2}} dt \frac{\partial \bar
H}{\partial \rho} &=& -\delta' I_{1,2}^\delta +
\frac{\lambda'}{2} I_{1,2}^{\lambda} = \label{integrals}\\
-\delta' \int_{w_{a,b}}^{w_s} dw \frac{w-w_s}{\dot{w}} &+& \frac{
\lambda'}{2} \int_{w_{a,b}}^{w_s} dw \frac{w^2-w_s^2}{\dot{w}},
\nonumber\eea where lower limits of integration for $I_1,I_2$ are
$w_a$ and $w_b$ correspondingly. For value of $\dot{w}$ one uses
the Eq. \ref{3factor} which makes the integrands in Eqs.
\ref{integrals} simple, and one gets

\bea \frac{\lambda}{2} I_1^{\delta}&=&\arcsin \left[\frac{-2 w_s + w_a + w_b}{w_b - w_a}\right] - \pi/2, \nonumber\\
\frac{\lambda}{2} I_1^{\lambda}&=& \sqrt{-(w_s-w_a)(w_s-w_b)} + \left(w_s + (w_a+w_b)/2\right) I_1^{\delta} ,\nonumber \\
\frac{\lambda}{2} I_2^{\delta} &=& -\arcsin \left[\frac{-2 w_s + w_a + w_b}{w_b - w_a}\right] - \pi/2, \\
\frac{\lambda}{2} I_2^{\lambda}&=& -\sqrt{-(w_s-w_a)(w_s-w_b)} +
\left(w_s + (w_a+w_b)/2\right) I_2^{\delta}\nonumber
 \eea

Therefore,

\bea {\cal P}_2 &=& \frac{I_2}{I_1} = \frac{-\delta' ( -\alpha -
\pi/2) + \frac{\lambda'}{2} [ -Q_s+W_s( -\alpha -
\pi/2)]}{-\delta' ( \alpha - \pi/2)
+ \frac{\lambda'}{2} [ Q_s+W_s( \alpha - \pi/2) ]  }, \nonumber\\
\alpha &=& \arcsin \left[\frac{-2 w_s + w_a + w_b}{w_b - w_a}\right], \label{explicit}\\
Q_s &=& \sqrt{-(w_s-w_a)(w_s-w_b)}, \quad W_s=w_s+(w_a+w_b)/2
\nonumber
 \eea

In the numerical example presented in Fig. \ref{fcapture},
$\delta'=-1$, $\lambda'=-\kappa=-1.5$; at the separatrix crossing
$\lambda_*=8.3863369$, $\delta_*= -3.0757753$, $h_s=0.3553544$. It
gives $w_a \approx -0.9239628$, $ w_b\approx0.30155167$,
$w_s\approx-0.4223149$.  The formula (\ref{explicit}) gives ${\cal
P}_2 \approx 86.998$, which perfectly corresponds to the numerical
result (87\%).


\begin{thebibliography}{99}
\bibitem{dynamical}
Note that the geometric change in the action is accompanied by the
nontrivial $\eps-$dependent {\em dynamical} change mentioned
above. It is usually the {\em dynamical} change in the action
which causes the {\em destruction of adiabatic invariance}.
\end{thebibliography}
\end{document}